\documentclass[twocolumn,epjc3]{svjour3}
\RequirePackage[T1]{fontenc}
\smartqed
\usepackage[dvipsnames]{xcolor}
\usepackage{amsfonts, amsmath, amssymb, newtxtext, newtxmath, slashed}
\usepackage[colorlinks,citecolor=magenta,urlcolor=cyan,linkcolor=blue]{hyperref}
\usepackage[utf8x]{inputenc} 
\usepackage[english]{babel}
\usepackage[export]{adjustbox}
\RequirePackage[numbers,sort&compress]{natbib}
\journalname{Eur. Phys. J. C}

\newcommand{\be}{\begin{equation}}
\newcommand{\ee}{\end{equation}}
\def\bsp#1\esp{\begin{split}#1\end{split}}

\begin{document}

\title{Collider imprint of vector-like leptons in light of anomalous magnetic moment and neutrino data}

\author{
  Parham Dehghani\,\thanksref{e1,add1} \and
  Mariana Frank\,\thanksref{e2,add1} \and
  Benjamin Fuks\,\thanksref{e3,add2}
}

\thankstext{e1}{\href{parham.dehghani@concordia.ca}{parham.dehghani@concordia.ca}}
\thankstext{e2}{\href{mariana.frank@concordia.ca}{mariana.frank@concordia.ca}}
\thankstext{e3}{\href{mailto:fuks@lpthe.jussieu.fr}{fuks@lpthe.jussieu.fr}}

\institute{
  Department of Physics, Concordia University, 7141 Sherbrooke St. West, Montreal, Quebec H4B 1R6, Canada\label{add1} \and\
  Laboratoire de Physique Théorique et Hautes Énergies (LPTHE), UMR 7589, Sorbonne Université et CNRS, 4 place Jussieu, 75252 Paris Cedex 05, France\label{add2}
}

\date{\today}
\maketitle

\begin{abstract}
We investigate the impact of incorporating vector-like leptons into the Standard Model, aiming to address longstanding puzzles related to the anomalous magnetic moments of the muon and electron while maintaining consistency with neutrino masses and mixings. We find that among the various representations of vector-like leptons permitted by the Standard Model gauge symmetry, only weak doublets and singlets offer satisfactory solutions, all associated with a significantly constrained parameter space. Our analysis delves into the associated parameter space, identifying representative benchmark scenarios suitable for collider studies. These setups yield a distinctive six-lepton signature whose associated signals can easily be distinguished from the Standard Model background, providing a clear signal indicative of new physics models featuring vector-like leptons. Our work hence sheds light on the potential implications of vector-like leptons in resolving discrepancies inherent to the Standard Model, while also offering insights into experimental avenues for further exploration.
\end{abstract}

%%%%%%%%%%%%%%%%%%%%%%%%%%%%%%%%%%%%%%%%%%%%%%%%%%%%%%%%%%%%%%%%%%%%%%%%%%%%%%
\section{Introduction}
\label{sec:intro}
%%%%%%%%%%%%%%%%%%%%%%%%%%%%%%%%%%%%%%%%%%%%%%%%%%%%%%%%%%%%%%%%%%%%%%%%%%%%%%
While the Standard Model (SM) has achieved remarkable success in describing fundamental particles and their interactions, its formulation includes important theoretical shortcomings. These have prompted extensive exploration of models Beyond the Standard Model (BSM), which often involve enlarging the particle content and/or gauge structure of the SM. Such extensions aim to address specific challenges inherent to the SM, including the existence of dark matter, the origin of neutrino masses, and the asymmetry between matter and antimatter in the universe. On the other hand, discrepancies between SM predictions and experimental measurements provide strong motivation for the development of new physics theories. 

One notable example of this tension is evident in the precise measurement of the muon's anomalous magnetic moment, denoted $a_{\mu}$, and the associated theoretical predictions. Combining data from experiments at Fermilab (FNAL) and Brookhaven National Laboratory (BNL), we get an experimental value given by~\cite{Muong-2:2021ojo,Muong-2:2006rrc, Muong-2:2023cdq},
\be
  a_\mu^{\rm exp}\,(\text{FNAL+BNL})=116~592~059\,(22) \times 10^{-11}\,(0.19\,\rm ppm)\,.
\ee
This measured value exhibits a significant deviation from the SM prediction,
\be
  a_\mu^{\rm SM}=116~591~810\,(43) \times 10^{-11}\,,
\ee
as evaluated by the Muon $g-2$ Theory Initiative in 2020~\cite{Aoyama:2020ynm}. However, the recent evaluation of the leading-order hadronic vacuum polarisation contribution to $a_\mu$ using lattice simulations suggests a reconciliation between the theoretical and experimental values for $a_\mu$~\cite{Borsanyi:2020mff, Ce:2022kxy, ExtendedTwistedMass:2022jpw, RBC:2023pvn, FermilabLatticeHPQCD:2023jof}. These lattice results nevertheless lead to discrepancies with the measurements of properties of $e^+e^-\to\pi^+\pi^-$ scattering~\cite{Wittig:2023pcl}, leaving hence the possibility open for new physics contributions. Thus, the $(g-2)_\mu$ puzzle continues to motivate the exploration of BSM avenues for restoring agreement between theory and data, as showcased for instance in~\cite{Jegerlehner:2009ry}.

On the other hand, there also exist tensions between measurements and SM predictions for the electron anomalous magnetic moment $a_e$. In the SM, $a_e$ computations include QED corrections up to the tenth order, and they rely on the measurement of the fine-structure constant using the recoil velocity and frequency of atoms that absorb a photon~\cite{Aoyama:2017uqe},
\be 
  a_e^{\rm SM} =1159652181.643\times 10^{-12}\,.
\ee
The most recent combined experimental measurements, determined in~\cite{Fan:2022eto}, yield
\be 
  a_e^{\rm exp}=115965218(12) \times 10^{-11}\,,
\ee
although there is currently a $5.5\sigma$ discrepancy between data obtained using
Rubidium-87~\cite{Morel:2020dww} and Cesium-133~\cite{Parker:2018} atoms,
\begin{equation}
 \Delta a_e \equiv \,a^{\text{exp}}_{e}-a^{\text{SM}}_{e} = 
    \left\{ \begin{array}{cc} 
       48(30) \times 10^{-14} & {\rm Rb}\\
      -88(36) \times 10^{-14} &{\rm Cs}
    \end{array} \right. ~.
\end{equation}
As in the case of the anomalous magnetic moment of the muon, this puzzle could be indicative of the presence of physics beyond the SM.

Several BSM extensions propose the existence of new exotic heavy leptons, notably vector-like leptons (VLLs) with identical left-handed and right-handed couplings akin to those originating from composite paradigms~\cite{Panico:2015jxa, Cacciapaglia:2020kgq, Cacciapaglia:2022zwt} or present in several neutrino mass models~\cite{Cai:2017mow}. Equipping particle physics theories with VLLs has moreover shown promises in addressing various challenges and anomalies. For instance, BSM setups with VLLs can encompass a dark matter candidate~\cite{Schwaller:2013hqa, Bahrami:2016has}, and they may contribute to resolving discrepancies in the muon's anomalous magnetic moment, hence potentially restoring agreement between theory and data~\cite{Poh:2017tfo, Crivellin:2018qmi, Hiller:2019mou, Hiller:2020fbu, Chun:2020uzw, De:2021crr, Escribano:2021css, Lee:2022nqz}. Additionally, VLLs could predict deviations from lepton universality in $B$-decays~\cite{Poh:2017tfo}, offer an explanation for the Cabibbo angle anomaly~\cite{Crivellin:2020ebi}, and align predictions of the $W$-boson mass with recent experimental measurements from the CDF collaboration~\cite{Lee:2022nqz}.

Experimental constraints on VLLs remain relatively weak and highly dependent on the specific model under consideration. These constraints primarily depend on the VLL's representation under the electroweak symmetry group and their couplings to SM leptons. For example, weak VLL doublets coupling to taus are constrained to be heavier than approximately 1~TeV~\cite{CMS:2019hsm, ATLAS:2023sbu}, while the limit drops to the range of 100-200~GeV in case of VLL singlets~\cite{CMS:2022nty}. Additionally, bounds from previous experiments at LEP are always valid, imposing a lower limit on the VLL mass of about 100~GeV, the exact value depending on the specific model details~\cite{ALEPH:1996akm, DELPHI:2004ili, L3:2001xsz, OPAL:2002wkp}. 

One crucial motivation for including VLLs in theoretical frameworks is their potential to extend the SM neutrino sector to accommodate a variety of experimental data. Neutrinos, among the most mysterious particles in the universe, play essential roles in understanding weak interactions, as well as cosmology, nuclear physics and astrophysics. While expected to be massless in the SM, neutrinos have been shown to oscillate between flavours, indicating that they have non-zero masses. While the neutrino absolute masses are not known, neutrino oscillations measured mass-square differences, found to be $m_{\nu_2}^2-m_{\nu_1}^2=7.5 \times 10^{-5}$ eV$^2$ and $m_{\nu_3}^2-m_{\nu_1}^2=\pm 2.5 \times 10^{-3}$ eV$^2$~\cite{deGouvea:2016qpx}. Additionally, cosmic surveys indirectly constrain the sum of neutrino masses to be at the sub-eV level, $\sum_j m_{\nu_j} <0.23$~eV at the 95\% confidence level~\cite{Loureiro:2018pdz}. The discovery of non-zero neutrino masses is probably among the most important particle physics results of the last two decades, as it suggests that the SM is incomplete. Thus any consistent model of BSM physics should be able to provide a mechanism and an explanation for neutrino masses and mixings.

In this work, we conduct a comprehensive analysis of models incorporating VLLs irrespective of their representation under the SM gauge symmetry, with a main focus on their consequences at colliders. We explore various BSM constructions with VLL singlets, doublets and triplets, and we probe their potential role in addressing fundamental physics puzzles such as dark matter, the muon and electron anomalous magnetic moments, and neutrino masses (all of these corresponding to limitations of the SM that could be related to the existence of an extended leptonic sector). Despite introducing only one or multiple VLL representations, we find no common explanation that could account for all these phenomena. While the idea of dark matter being associated with a vector-like neutrino has been extensively discussed in prior literature~\cite{Jegerlehner:2009ry}, our emphasis lies on options that provide explanations for discrepancies between theoretical predictions and experimental measurements in the context of the lepton anomalous magnetic moments, and that are suitable with respect to neutrino mass and mixing constraints. Additionally, we ensure that our solutions are compatible with electroweak precision data, and finally we investigate their implications at collider experiments.

The connection between the anomalous magnetic moment of the muon and electron, along with the presence of vector-like leptons in the theory, has been previously studied in several studies. In~\cite{Kannike:2011ng}, old constraints originating from the puzzle associated with the anomalous magnetic moment of the muon were used to define interesting collider signals. Furthermore, Ref.~\cite{Arcadi:2021cwg} recently analysed the connection between $(g-2)_\mu$, dark matter and $B$-meson anomalies independently of collider studies, in complementarity with explorations of the impact of vector-like leptons on $(g-2)_\mu$ and its correlations with the properties of the Higgs sector, and either leptonic flavour violation~\cite{Poh:2017tfo}, a dark matter candidate~\cite{Lu:2021vcp}, the measurement of the $W$-boson mass by the CDF experiment~\cite{CDF:2022hxs, deGiorgi:2022xhr, Lee:2022nqz}, or even in an effective field theory interpretation~\cite{DelleRose:2022ygn}. However, none of these works deeply investigated the consequences on potential VLL signals at colliders by means of a state-of-the-art simulation tool chain. Non-minimal fermion-scalar TeV-scale extensions of the SM were also explored, but again without examining any VLL collider implication~\cite{Guedes:2022cfy}. In addition, a joint explanation to observed deviations in both the electron and muon anomalous magnetic moments was discussed in~\cite{Craig:2021ksw} through the identification of a hidden symmetry responsible for a ``magic zero'' implying that the leading contribution to the muon $g-2$ vanishes upon integrating out weak doublet and singlet of vector-like fermions. Non-minimal model studies also include setups in which the SM is augmented not only by vector-like leptons, but also by additional gauge symmetries~\cite{Kawamura:2022fhm, Zhou:2022cql} or an extended scalar sector~\cite{Chun:2020uzw, Raju:2022zlv, Frank:2020smf, DeJesus:2020yqx, Bharadwaj:2021tgp, Chakrabarty:2020jro}, as well as in warped space constructions~\cite{Megias:2017dzd}.

In comparison, our work aims to bring two new important features complementing previous explorations. First, our analysis includes the most up-to-date constraints from both BNL and FNAL experiments for the anomalous magnetic moment of the muon, and second, it includes a comprehensive collider analysis for promising benchmarks which are consistent with \textit{both} satisfying constraints from low energy data, neutrino masses and mixings, and anomalous moment measurements of both the muon and the electron.

The rest of this study is organised as follows. In section~\ref{sec:model} we introduce our theoretical framework and write down the VLL Lagrangian that we consider, we present the experimental constraints included in our analysis, and we introduce the software tools that we use to conduct a comprehensive parameter space scan. Section~\ref{sec:scanresults} is dedicated to a presentation of our results, with a special focus on the anomalous magnetic moments of the electron and muon in section~\ref{subsec:g-2}, on neutrino data in section~\ref{subsec:neutrinos}, and on electroweak precision tests in section~\ref{subsec:oblique}. In section \ref{sec:collider} we select representative benchmark scenarios and determine their collider signatures. Finally, we summarise our findings and conclude in section~\ref{sec:conclusion}.

%%%%%%%%%%%%%%%%%%%%%%%%%%%%%%%%%%%%%%%%%%%%%%%%%%%%%%%%%%%%%%%%%%%%%%%%%%%%%
\section{Theoretical Framework}
\label{sec:model}
%%%%%%%%%%%%%%%%%%%%%%%%%%%%%%%%%%%%%%%%%%%%%%%%%%%%%%%%%%%%%%%%%%%%%%%%%%%%%%
Unlike SM-like chiral fermions, whose left-handed and right-handed components transform differently under the SM gauge symmetry $SU(3)_C\otimes SU(2)_L \otimes U(1)_Y$, vector-like fermions exhibit identical properties for both their chiralities. However, the specific choice of representation for these fermions is not unique. In models featuring VLLs, building a consistent, renormalisable, and anomaly-free (as automatically guaranteed by the addition of vector-like fermions) extension of the SM imposes constraints on the possible choices for their $SU(2)_L$ and hypercharge quantum numbers, especially once we consider only gauge-covariant lepton multiplets which can mix with the ordinary leptons through renormalisable couplings. In other words, we will only consider interactions with ordinary fermions through Yukawa couplings to the SM scalar~\cite{Aguilar-Saavedra:2013qpa}.

\begin{table*}
  \begin{center}\renewcommand{\arraystretch}{1.2}
  \begin{tabular*}{.95\textwidth}{@{\extracolsep{\fill}} c| cccccc}
    &${\cal S}_1$ &${\cal S}_2$ &${\cal D}_1$ & ${\cal D}_2$  & ${\cal X}_1$  &${\cal X}_2$\\
	\hline
    &&&&&\\[-.4cm]
	Multiplets &$E^-$ &$N^0$ & $\begin{pmatrix} {\cal D}_1^0 \\ {\cal D}_1^- \end{pmatrix} $ & $\begin{pmatrix} {\cal D}_2^- \\ {\cal D}_2^{--}\end{pmatrix}$  &$ \begin{pmatrix} \frac{1}{\sqrt{2}} {\cal X}_1^0 & {\cal X}_1^+\\  {\cal X}_1^- & -\frac{1}{\sqrt{2}} {\cal X}_1^0 \end{pmatrix}$ & $\begin{pmatrix} \frac{1}{\sqrt{2}} {\cal X}_2^- & {\cal X}_2^0\\  {\cal X}_2^{--} & -\frac{1}{\sqrt{2}} {\cal X}_2^- \end{pmatrix}$ \\[.6cm]
    \hline
    $SU(2)_L$ & $\mathbf{1}$ & $\mathbf{1}$ & $\mathbf{2}$ & $\mathbf{2}$  & $\mathbf{3}$ & $\mathbf{3}$ \\
    $U(1)_Y$       & $-1$ & 0 & $-1/2$ & $-3/2$  &$0$ &$-1$ \\
  \end{tabular*}
  \caption{\label{tab:VLrepresentations} Possible VLL representations leading to a renormalisable BSM theory. We consider weak singlets (${\cal S}_1$ and ${\cal S}_2$), doublets (${\cal D}_1$,  ${\cal D}_2$), and triplets (${\cal X}_1$ and ${\cal X}_2$) with different hypercharge quantum number assignments. In the second line of the table, we provide the components of the various weak multiplets considered (with their electric charge explicitly indicated), while the $SU(2)_L\times U(1)_Y$ quantum numbers are given in the last lines of the table.}
  \end{center}
\end{table*}

The list of acceptable choices, along with our notation for the fields and their weak multiplet components, is provided in Table~\ref{tab:VLrepresentations}. The first two singlet possibilities, ${\cal S}_1$ and ${\cal S}_2$, correspond to VLLs with quantum numbers matching those of a right-handed charged lepton $E$ or neutrino $N$ (considered as a Dirac fermion in the following), respectively. In addition, VLL doublets can be either SM-like or possess different quantum numbers (${\cal D}_1$, and ${\cal D}_2$). In the cases of the ${\cal D}_1$ state, the doublet component fields correspond to a neutral and singly-charged lepton, while for ${\cal D}_2$ they include a singly-charged and a doubly-charged state.  Similarly, the first triplet option, ${\cal X}_1$, includes only neutral and singly-charged VLLs and is considered as a Dirac fermion in the following, whereas the second possibility, ${\cal X}_2$, is more exotic and features a neutral, a singly-charged and a doubly-charged component.
 
To address the anomalous magnetic moment of both the electron and the muon as well as other puzzles of the SM such as neutrino masses and mixings, we construct a theoretical framework that incorporates one or more of the considered VLL representations, each endowed with a generation index. In other words, we enforce that a given VLL species primarily interacts with the corresponding generation of ordinary leptons, and next neglect any inter-generational coupling. This approach, distinct from introducing VLL fields with inter-generational couplings, allows us to analyse the impact of the model on the anomalous magnetic moments of the electron and muon separately, in a way that avoids inducing lepton-flavour-violating (LFV) decays. While it is arguable that inter-generational mixing is generated nevertheless through quantum corrections, LFV constraints enforce it to be small. The VLL collider signatures of our setup and their corresponding rates (including production cross sections and relevant branching ratios) depend on how the different VLLs mix with ordinary leptons, which stems from their Yukawa interaction with the Higgs field $\Phi$, as well as their mass terms. The corresponding Lagrangian contributions, with generational indices understood, are given by
\be\label{eq:lag1}\bsp 
  & {\cal L}_\mathrm{Y} = - \bigg[ 
      \mathbf{Y_{e}}\, {\bar L} \Phi e 
    + \mathbf{Y_{\nu}}\, ({\bar L} \cdot \Phi^\dag) \nu   
    + \mathbf{\hat{Y}_{\mathcal{S}_1}}\, {\bar L} \Phi {\cal S}_1\\
    &\quad
    + \mathbf{\hat{Y}_{\mathcal{S}_2}}\, ({\bar L} \!\cdot\! \Phi^\dag) {\cal S}_2
    + \mathbf{\hat{Y}_{\mathcal{D}_1}}\, {\bar {\cal D}}_1 \Phi e
    + \mathbf{\hat{Y}_{\mathcal{D}_2}}\, ({\bar {\cal D}}_{2} \!\cdot\! \Phi^\dag) e\\
    &\quad
    + \mathbf{\hat{Y}_{\mathcal{X}_1}}\, {\bar L} {\cal X}_{1}  \Phi
    + \mathbf{\hat{Y}_{\mathcal{X}_2}}\, \Phi \!\cdot\! ({\cal X}_{2} L)
    + \mathrm{H.c.} \bigg]  \\
    &\quad
    - \sum_{i=1}^2 \bigg[ M_{\mathcal{S}_i}\, {\bar {\cal S}}_{i} {\cal S}_{i}+ M_{\mathcal{D}_i}\, {\bar {\cal D}}_{i} {\cal D}_{i}
    + M_{\mathcal{X}_i}\, {\bar {\cal X}}_{i} {\cal X}_{i}\bigg]  \,.
\esp\ee
Here, $L$, $e$ and $\nu$ represent the SM doublet and singlet of ordinary leptons and neutrinos\footnote{While strictly speaking there is no right-handed neutrino $\nu\equiv\nu_R$ in the SM, we consider a SM sector containing such a state, that is thus differently tagged from the VLL representations studied. However, all bilinear terms involving a right-handed neutrino have been ignored for simplicity, as they do not impact our ability to find scenarios compatible with all constraints considered in this work and the associated collider phenomenology.}, respectively, and the explicit invariant products appearing in the Lagrangian denote $SU(2)$-invariant products of two fields lying in the fundamental or anti-fundamental representation of $SU(2)_L$. The different parameters appearing in the Lagrangian ${\cal L}_\mathrm{Y}$ include the $3\times 3$ SM Yukawa matrices in the flavour space $\mathbf{Y_{e}}$, $\mathbf{Y_{\nu}}$, $\mathbf{\hat{Y}_{\mathcal{S}_1}}$, $\mathbf{\hat{Y}_{\mathcal{S}_2}}$, $\mathbf{\hat{Y}_{\mathcal{D}_1}}$, $\mathbf{\hat{Y}_{\mathcal{D}_2}}$, $\mathbf{\hat{Y}_{\mathcal{X}_1}}$ and $\mathbf{\hat{Y}_{\mathcal{X}_2}}$, along with the mass parameters associated with the VLLs $M_{\mathcal{S}_1}$, $M_{\mathcal{S}_2}$, $M_{\mathcal{D}_1}$, $M_{\mathcal{D}_2}$, $M_{\mathcal{X}_1}$ and $M_{\mathcal{X}_2}$. All these matrices are imposed to be flavour diagonal, which ensures that a vector-like representation of a given generation only interacts with the SM lepton of the same generation. Furthermore, terms involving different VLL species have all been omitted for now, and will be discussed below.

Constructing a new physics framework suitable for explaining dark matter with the desired properties singles out the singlet representation ${\cal S}_2$. This representation is the only one that includes an electrically neutral state that can be stabilised through an additional $\mathbb{Z}_2$ parity, once we enforce that all SM states are $\mathbb{Z}_2$-even and $N^0$ is $\mathbb{Z}_2$-odd. However, this configuration prevents neutrinos from mixing and acquiring mass, and has no impact on predictions for the anomalous magnetic moments of the electron and the muon. Since the focus of the present work does not encompass dark matter, we do not consider this possibility. Instead, we utilise the possibility of supplementing the SM with an electrically-neutral vector-like singlet ${\cal S}_2$ along with the SM-like doublet ${\cal D}_1$ to generate mixing between all neutral leptons, account for the observed neutrino masses, and satisfy all experimental bounds based on neutrino oscillations. Furthermore, our numerical analysis indicates that meaningful and acceptable contributions to the anomalous magnetic moments of the electron and the muon can additionally be obtained by including both the SM-like doublet ${\cal D}_1$ and the charged vector-like singlet ${\cal S}_1$ in the model, which mix with the SM doublets of ordinary leptons $L$. 

While embedding the remaining representations shown in Table~\ref{tab:VLrepresentations} into the model could also yield acceptable solutions for neutrino mass generation and resolve the puzzles related to the anomalous magnetic moments of the muon and the electron, this option quickly becomes highly non-minimal and escalates in a non-necessary complexity for an initial exploration of such BSM models. Therefore, we limit our study to an extension of the field content of the SM in which we incorporate three generations of ${\cal S}_1$, ${\cal S}_2$ and ${\cal D}_1$ VLLs, deferring the analysis of other options to future work. The incorporation of the three sets of VLL species ${\cal S}_1$, ${\cal S}_2$ and ${\cal D}_1$ into the model necessitates supplementing the Lagrangian by extra contributions involving several of the VLL species considered. The terms allowed by the SM gauge symmetry read
\be \label{eq:lag2}\bsp
  {\cal L}_\mathrm{VLL} =&\ \bigg[
       \big({\bar {\cal D}}_1 \Phi\big) \big( \lambda_L^{(1)} P_L + \lambda^{(1)}_R P_R \big) {\cal S}_1 \\
    &\quad
    + \big({\bar {\cal D}}_1 \cdot \Phi^\dag\big) \big( \lambda_L^{(2)} P_L + \lambda^{(2)}_R P_R \big) {\cal S}_2 
    + \mathrm{H.c.} \bigg]  \,.
\esp\ee
In this expression, $P_L$ and $P_R$ represent the left-handed and right-handed chirality projectors, respectively, while the couplings $\lambda_{L,R}^{(1,2)}$ are matrices in the flavour space. Once again, these matrices are imposed to be diagonal, which automatically ensures that each additional VLL is connected to a single generation. 

\begin{table}
  \begin{center}\renewcommand{\arraystretch}{1.4}
  \begin{tabular}{ccc}
	Observable & Constraints\\
	\hline
	$m_h$ & [122, 128] GeV\\
	BR$(h\rightarrow\mu^+\mu^-)$ (strength) & $[0.51, 1.87]$\\
	BR$(h\rightarrow \gamma \gamma)$ (strength) & $[0.96, 1.24]$\\
	BR$(B^0_s \rightarrow \mu^{+}\mu^{-})$&$[1.1, 6.4] \times 10^{-9}$\\
	$\mathrm{BR}(B^0 \rightarrow \chi_s\gamma)$ & $[2.99, 3.87] \times 10^{-4}$ \\
	$\frac{\mathrm{BR}(B \rightarrow \tau \nu_{\tau})}{\mathrm{BR}_{{\rm SM}}(B \rightarrow \tau \nu_{\tau})}$ & [0.15, 2.41] \\
    $m_{\nu_1}+m_{\nu_2}+m_{\nu_3}$ & $\leq 0.26$ eV\\
    Neutrino masses& See section~\ref{subsec:neutrinos}
  \end{tabular}
  \caption{Experimental bounds imposed on the SM sector during our parameter space scanning procedure.}\label{tab:constraints}
  \end{center}
\end{table}

Subsequently, we proceed to demonstrate, by means of a scan over the model's free parameters, that there exist benchmark scenarios  allowing a broad range of low-energy constraints to be satisfied. Moreover, we show that these choices fulfil requirements from neutrino oscillation data and measurements of the anomalous magnetic moment of the electron and the muon. More precisely, we include constraints from the results of recent LHC searches for VLLs, which require us to impose lower bounds on the mass of any charged VLL state~\cite{CMS:2019hsm, CMS:2022nty, ATLAS:2023sbu}. Additionally, we ensure that any scenario retained in the scan features a Higgs boson with mass and decay properties consistent with observational data~\cite{ATLAS:2020fzp, ATLAS:2023oaq, CMS:2020xrn, CMS:2020xwi, Workman:2022ynf}, up to potential parametric uncertainties arising from the new physics sector. Furthermore, we impose constraints from rare $B$-hadron decays~\cite{HFLAV:2022pwe, LHCb:2013ghj, LHCb:2012skj} and neutrino cosmological data \cite{Loureiro:2018pdz}. These constraints are compiled in Table~\ref{tab:constraints}. 

\begin{table}
  \begin{center}\renewcommand{\arraystretch}{1.3}\setlength{\tabcolsep}{20pt}
  \begin{tabular}{cc}
    Parameter& Scanned range\\
	\hline
	$|\mathbf{\hat{Y}_{\mathcal{D}_1}}|$ & $[10^{-5}, 1]$\\ 
    $|\mathbf{\hat{Y}_{\mathcal{S}_1}}|$ & $[10^{-5}, 1]$  \\ 
    $|\mathbf{\hat{Y}_{\mathcal{S}_2}}|$ & $[10^{-7}, 10^{-3}]$\\  
    $|\mathbf{Y_{\nu}}|$ & $[10^{-15}, 10^{-10}]$\\  
	$|\lambda_{L}^{(1)}|\,,\ |\lambda_{R}^{(1)}|$ & $[10^{-5}, 1]$\\
    $|\lambda_{L}^{(2)}|\,,\  |\lambda_{R}^{(2)}|$ & $[10^{-7}, 10^{-3}]$\\
    $M_{\mathcal{S}_1}$ & [100, 500] GeV\\
    $M_{\mathcal{S}_2}$& [100, 1000]\\
    $M_{\mathcal{D}_1}$ & [100, 300] GeV\\   	
  \end{tabular}
  \caption{Ranges over which the model's free parameters are allowed to vary. These ranges are uniformly applied to each component of the diagonal Yukawa coupling and mass matrices in the flavour space.} \label{tab:parameterscan}
  \end{center}
\end{table}

Furthermore, it is important to note that VLL mixings with the SM lepton fields entail modifications of the couplings of the $W$ and $Z$ bosons with the SM-like leptons $\ell$. The relevant neutral-current Lagrangian terms can be expressed as follows, considering two SM leptons with flavour indices $i$ and $j$ (where $i,j=1,2,3$ correspond to the electron, muon and tau, respectively):
\begin{equation}\label{ZCoupling}\begin{split}
  & {\cal L}_{Z\ell\ell} = 
    - i  {\bar \ell}_i \bigg[L_\ell \sum_{k=1}^{6} U^{L,\,jk\,*}_\ell U^{L,\,ik}_\ell\\
    &\hspace{2.25cm}  + R_\ell \sum_{k=1}^{3}U^{L,\,j(6+k)\,*}_\ell U^{L,\,i(6+k)}_\ell \bigg]P_L\slashed{Z}\ell_j \\  
    &\ \  - i {\bar \ell}_i \bigg[ R_\ell  \sum_{k=1}^{3} \Big(U_\ell^{R,\,j k\,*} U_\ell^{R,\,i k} + U^{R,\,j(6+k)\,*}_\ell U^{R,\,i(6+k)}_\ell \Big) \\
    &\hspace{2.25cm} + L_\ell \sum_{k=1}^{3}U^{R,\,j(3+k)\,*}_\ell U^{R,\,i(3+k)}_\ell \bigg]P_R\slashed{Z} \ell_j\,,
\end{split}\end{equation}
where $L_\ell = g/c_W (-1/2 + s_W^2)$ and $R_\ell = g/c_W s_W^2$ represent the SM $Z$-boson couplings to left-handed and right-handed leptons. Here, $s_W$ and $c_W$ denote the sine and cosine of the electroweak mixing angle, respectively, and $g$ stands for the weak coupling constant. This interaction Lagrangian explicitly depends on the lepton mixing matrices $U^L_\ell$ and $U^R_\ell$ that diagonalise the mass matrix
\begin{equation}\renewcommand{\arraystretch}{1.2}
  m_{\ell} = \begin{pmatrix}
    \frac{v}{\sqrt{2}} \mathbf{Y_{e}}^{\!\!T}  &0 &\frac{v}{\sqrt{2}} \mathbf{\hat{Y}_{\mathcal{S}_1}} \\ 
    \frac{v}{\sqrt{2}} \mathbf{\hat{Y}_{\mathcal{D}_1}}  &M_{\mathcal{D}_1} &\frac{v}{\sqrt{2}} \lambda_R^{(1)}\\ 
    0 &\frac{v}{\sqrt{2}} \lambda_L^{(1)\,T}  &M_{\mathcal{S}_1}\end{pmatrix} \,, 
\end{equation}
such that
\begin{equation} 
  {\rm diag}(m_e, m_\mu, \ldots) = U^{L\,*}_\ell\, m_{\ell}\, U^{R\,\dagger}_\ell \,.
\end{equation}
In this expression, $v$ represents the vacuum expectation value of the SM Higgs doublet. Due to the significant differences in the orders of magnitude among the parameters present in the mass matrix, inter-generational mixing for the ordinary leptons is found to be negligible. Consequently, the couplings in Eq.~\eqref{ZCoupling} that involve ordinary leptons can be approximated to those of the SM. For example, in the case of the $Z$-boson coupling to the electron, we obtain:
\begin{align}
 {\cal L}_{Ze e} \approx -i {\bar e} \Big[ L_\ell P_L + R_\ell P_R\Big]\slashed{Z} e \,.
\end{align} 
Thus, the model examined in this study aligns with the established properties of the $Z$ boson. Similarly, the charged-current Lagrangian involving $W$-boson couplings to an ordinary charged lepton $\ell_j$  and neutrino $\nu_i$ (for $i=1,2,3$ being a flavour index) is given by
\begin{equation}\begin{split}
 {\cal L}_{W \nu_i \ell_j}=&\ 
   \frac{ig}{\sqrt{2}} \bar\nu_i \slashed{W}  \sum_{k=1}^{3} \bigg[ \\
     &\quad \Big(U_\ell^{L,\,j(3 + k)\,*} U_\nu^{L,\,i (3 + k)} - U_\ell^{L,\,j k\,*} U_\nu^{L,\,i k}\Big) P_L \\
     &\quad +  U_\ell^{R,\,j(3 + k)\,*} U_\nu^{R,\,i (3 + k)} P_R\bigg] \ell_j + \mathrm{H.c.} 
\end{split} \end{equation}
This Lagrangian depends both on the matrices $U^L_\ell$ and $U^R_\ell$ diagonalising the charged lepton sector, and on the matrices $U_{\nu}^L$ and $U_{\nu}^R$ that diagonalise the neutrino mass matrix,
\begin{equation} 
m_{\nu} = \begin{pmatrix}
\frac{1}{\sqrt{2}} v \mathbf{Y_{\nu}}  &0 &\frac{1}{\sqrt{2}} v \mathbf{\hat{Y}_{\mathcal{S}_2}} \\ 
0 &M_{\mathcal{D}_1} &\frac{1}{\sqrt{2}} v \lambda_R^{(2)} \\ 
0 &0 &M_{\mathcal{S}_2}\end{pmatrix}\end{equation}
such that
\begin{equation} 
{\rm diag}(m_{\nu_1}, m_{\nu_2}, \ldots) = U_{\nu}^{L\,*} m_{\nu} U_{\nu}^{R\,\dagger}\,.
\end{equation}
Once again, by virtue of the typical values of the different parameters entering that mass matrix, the $W$-boson couplings to SM leptons and neutrinos can be approximated as in the SM,
\begin{align}
 {\cal L}_{W \nu_i \ell_j} \approx -\frac{ig}{\sqrt{2}}  {\bar \nu_i}\slashed{W} P_L \ell_j \,,
 \end{align} 
such that the model examined in this work does not challenge measured $W$-boson couplings.
 
To explore the model's parameter space, we make use of the {\tt SARAH} package (version 4.15.1)~\cite{Staub:2013tta, Goodsell:2017pdq} along with an in-house implementation of the model described above. Our parameter space scan involves a randomised sampling over the model parameters within the ranges specified in Table~\ref{tab:parameterscan}. To ensure a thorough probe of diverse regions across the parameter space, we employ logarithmic samplings for all Yukawa couplings, as the lower part of the ranges are typically of greater interest and more likely to satisfy existing experimental constraints. Moreover, we bias our results towards light or moderately light VLL states, a choice that enhances the prospect of detecting these particles at the LHC or future high-energy colliders, and therefore testing our BSM construction in the near future. To evaluate the implications of each parameter set, we utilise the {\tt SPheno} package (version 4.0.5)~\cite{Porod:2011nf} to compute the associated mass spectrum, decay tables, and relevant information for the experimental observables under investigation.

We next proceed to evaluate, throughout the scan, how each benchmark scenario selected could potentially address discrepancies between theoretical predictions and experimental measurements regarding the anomalous magnetic moment of the muon $a_{\mu}$. Specifically, we assess the extent to which each scenario predicts a value of $a_{\mu}$ with up to a $3\sigma$ deviation from the combined result of experiments at BNL and FNAL~\cite{Muong-2:2023cdq}, 
\be\label{muon-g-2}
    \Delta a_{\mu}\,\equiv\,a^{\text{exp}}_{\mu}-a^{\text{SM}}_{\mu}\,=\,[\,105,\,393\,]\times10^{-11}\,.
\ee
Moreover, we also assess whether the selected scenarios can yield large deviations from the SM regarding the anomalous magnetic moment of the electron. In this case, we consider two distinct sets of measurements involving Cs and Rb atoms~\cite{Fan:2022eto}, and allow our predictions to deviate by 3$\sigma$ from Cs measurements and 2$\sigma$ from Rb measurements as
\be\label{electron-g-2}\bsp
  \Delta a^{\text{Cs}}_{e}\,\equiv & \,a^{\text{exp}}_{e}-a^{\text{SM}}_{e}\,=\,[\,-168,\,-108\,]\times10^{-14}\,,\\
   \Delta a^{\text{Rb}}_{e}\,\equiv & \,a^{\text{exp}}_{e}-a^{\text{SM}}_{e}\,=\,[\,2,\,66\,]\times10^{-14}\quad \,.
\esp\ee
There is some controversy in the literature regarding whether the anomalous magnetic moment of the muon constitutes a signal of physics beyond the SM. Recent lattice simulations of the hadronic vacuum polarisation contribution to $a_\mu$ have suggested a reconciliation between theoretical predictions and experimental measurements~\cite{Borsanyi:2020mff, Ce:2022kxy, ExtendedTwistedMass:2022jpw, RBC:2023pvn, FermilabLatticeHPQCD:2023jof}, while also revealing discrepancies in the properties of $e^+e^- \to \pi^+\pi^-$ scattering~\cite{Wittig:2023pcl}. As a result, while the likelihood that there are large new physics contributions to $a_\mu$ has diminished, some unresolved issues remain and continue to motivate BSM explorations in this area. This leaves open the question of whether the deviation in Eq.~\eqref{muon-g-2} would still be considered  a sign of new physics affecting hadronic vacuum polarisation. If this interpretation holds, it is important to remember that such contributions generally worsen electroweak precision fits. Conversely, models which predict smaller contributions to $a_\mu$ may become increasingly attractive in light of the recent lattice findings (and a negative answer to the above question).  For an up-to-date discussion, see \cite{Aoyama:2020ynm}.
%%%%%%%%%%%%%%%%%%%%%%%%%%%%%%%%%%%%%%%%%%%%%%%%%%%%%%%%%%%%%%%%%%%%%%%%%%%%%
\section{Scan Results}
\label{sec:scanresults}
%%%%%%%%%%%%%%%%%%%%%%%%%%%%%%%%%%%%%%%%%%%%%%%%%%%%%%%%%%%%%%%%%%%%%%%%%%%%%

After conducting a random scan across the parameter space of the model as defined in Table~\ref{tab:parameterscan}, we have identified 328 solutions that satisfy all bounds listed in Table~\ref{tab:constraints} along with restrictions on VLL masses and milder constraints originating from the anomalous magnetic moments of the muon and the electron to orient the scan towards the physics case studied. Many of these solutions hence exhibit significant deviations from the SM predictions for the anomalous magnetic moments of both the muon and the electron, potentially offering a pathway to reconciling theory with data. Specifically, these solutions feature $\Delta{a}_{\mu}$ values reaching up to $3\sigma$, as defined by Eq.~\eqref{muon-g-2}, and $\Delta{a}^{\rm Rb}_{e}$ values showing positive deviations of up to $2\sigma$, aligning with the results obtained from measurements carried out from Rubidium atoms and presented in Eq.~\eqref{electron-g-2}. However, none of the obtained solutions predicts a negative deviation for $a_e$ such that our model cannot explain $a_e$ values as extracted from measurements of properties of Cesium atoms measurements. In the following, we investigate these findings in detail, first relative to the anomalous magnetic moments of the muon and the electron in section~\ref{subsec:g-2}, and next in the context of neutrino data in section~\ref{subsec:neutrinos}. Finally, we verify in section~\ref{subsec:oblique} that adding multiplets of VLLs to the theory does not impact electroweak precision tests.

%%%%%%%%%%%%%%%%%%%%%%%%%%%%%%%%
\subsection{Anomalous magnetic moments of the muon and the electron}
\label{subsec:g-2}
%%%%%%%%%%%%%%%%%%%%%%%%%%%%%%%%%%

\begin{figure*}
  \centering 
  \includegraphics[width=\textwidth]{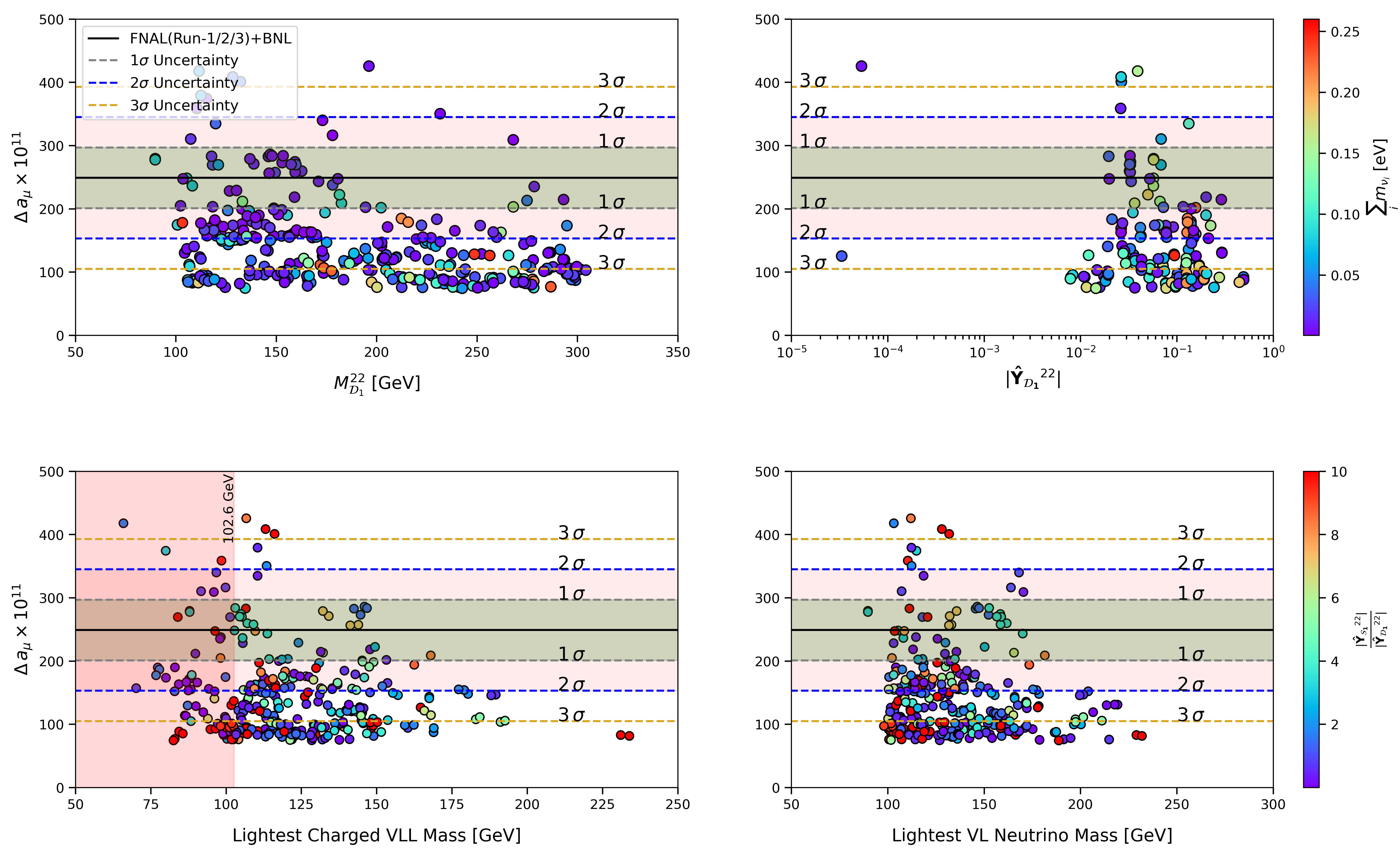}
  \caption{Predictions for the new physics contributions to the anomalous magnetic moment of the muon and their dependence on the second generation vector-like mass $M^{22}_{\mathcal{D}_1}$ (top left), the Yukawa coupling of the second-generation VLL doublet with the ordinary lepton singlet $|\mathbf{\hat{Y}_{\mathcal{D}_1}}^{22}|$ (top right) and the masses of the lightest vector-like charged lepton (bottom left) and neutrino (bottom right). Each point represents a scenario compliant with all constraints imposed during the scan, and the colour code represents either the sum of the masses of the three lightest neutrinos (top row) or the ratio of the second-generation Yukawa couplings of VLL doublet and singlet (bottom row).}\label{fig:1}
\end{figure*}

In figure~\ref{fig:1} we present various properties of the 328 scenarios identified in our scan. We display the dependence of the new physics contributions $\Delta{a}_{\mu}$ to the anomalous magnetic moment of the muon on the second-generation vector-like mass parameter $M^{22}_{\mathcal{D}_1}$ (top left panel), on the absolute value of the second-generation lepton-VLL Yukawa coupling $|\mathbf{\hat{Y}_{\mathcal{D}_1}}^{22}|$ (top right panel), as well as on the mass of the lightest charged VLL state (bottom left panel) and neutral VLL state (bottom right panel). We observe that the constraints enforced lead to relatively large values for the second-generation Yukawa coupling $|\mathbf{\hat{Y}_{\mathcal{D}_1}}^{22}|$, typically ranging from 0.01 and 0.5. Additionally, the colour coding in the bottom row indicates that the Yukawa coupling $|\mathbf{\hat{Y}_{\mathcal{S}_1}}^{22}|$ relating the second-generation VLL singlet $\mathcal{S}_1$ and the SM weak doublet $L$ must be even larger (by a factor of a few). On the other hand, the dependence of $\Delta{a}_{\mu}$ on the doublet VLL mass $M^{22}_{\mathcal{D}_1}$ is moderate, with values spanning the entire scanned range, like the one on the masses of the lightest vector-like states. 

\begin{figure*}
  \centering 
  \includegraphics[width=\textwidth]{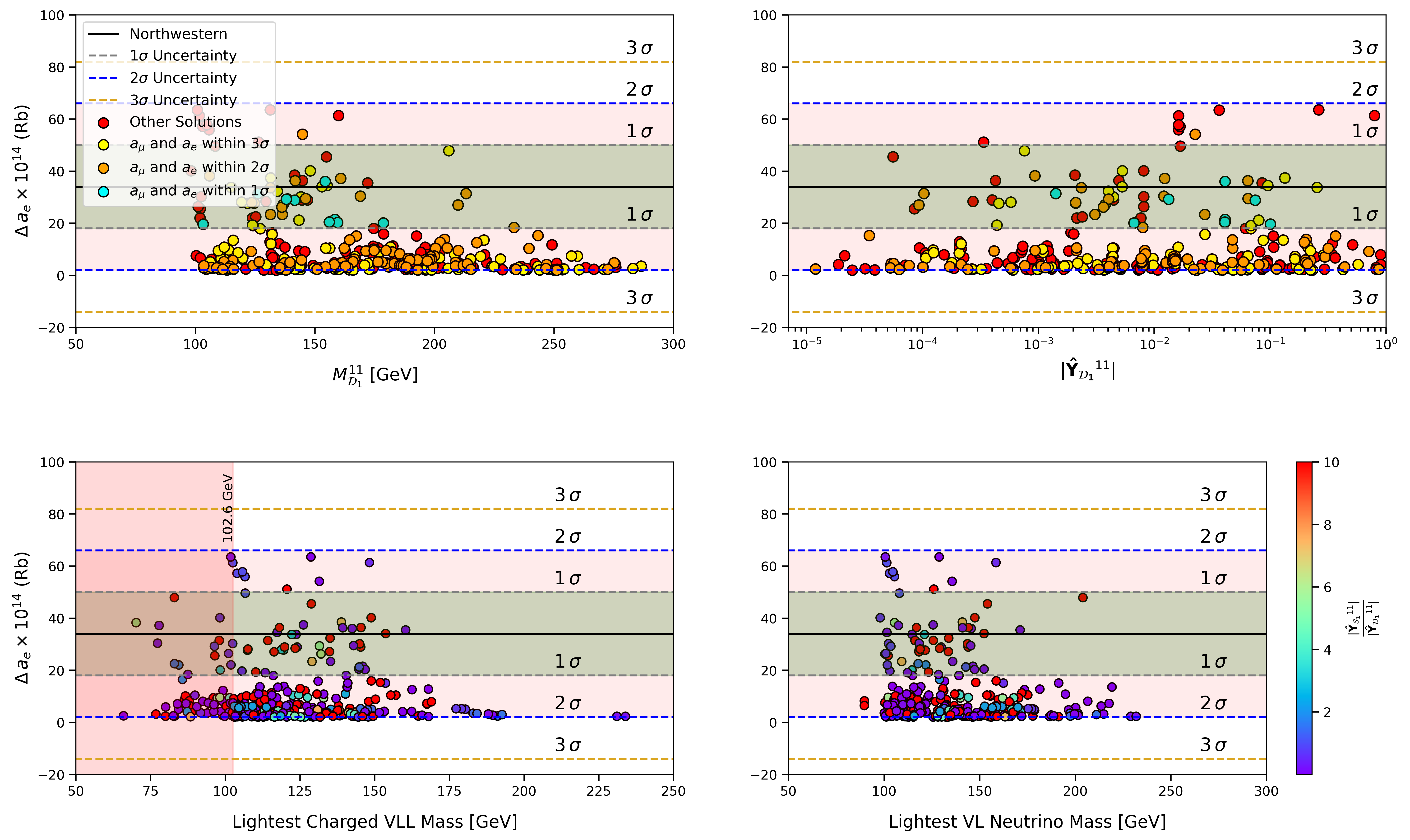}
  \caption{Predictions for the new physics contributions to the anomalous magnetic moment of the electron and their dependence on the first-generation vector-like mass parameter $M^{11}_{\mathcal{D}_1}$ (top left), the Yukawa coupling of the first-generation VLL doublet with the ordinary electron singlet $|\mathbf{\hat{Y}_{\mathcal{D}_1}}^{11}|$ (top right), the masses of the lightest vector-like charged lepton (bottom left) and neutrino (bottom right). Each point represents a scenario compliant with all constraints imposed during the scan, and the colour code either represents how scenarios can simultaneously provide an explanation for the two anomalous magnetic moments considered (top row), or it highlights the ratio of the first-generation Yukawa couplings of VLL doublet and singlet (bottom row). \label{fig:2}}
\end{figure*}

The observed pattern of large Yukawa couplings results in a complex mixing between the VLL doublet state $\mathcal{D}_1$, the VLL singlet state $\mathcal{S}_1$, and the left-handed and right-handed muon fields, while not spoiling observations and ensuring large values for new physics contributions to the anomalous magnetic moment of the muon. Subsequently, this provides sufficient flexibility to adjust the other input parameters of the model to achieve agreement with data in the neutrino sector or feature deviations in the anomalous magnetic moment of the electron that are significant enough to provide an explanation for its observed value. This is exemplified below for the anomalous magnetic moment of the electron, as well as in figure~\ref{fig:1} by the values obtained for the sum of the masses of the three lighter neutrinos (represented through the colour code in the top row of the figure) and in section~\ref{subsec:neutrinos}. Upon examination of the bottom row of figure~\ref{fig:1}, it becomes evident that a substantial portion of the solutions featuring a $3\sigma$ deviation from the SM predictions for $a_\mu$ also feature a VLL spectrum where the lightest state is predominantly singlet-like (as $|\mathbf{\hat{Y}_{\mathcal{D}_1}}^{22}| < |\mathbf{\hat{Y}_{\mathcal{S}_1}}^{22}|$). Nonetheless, we have verified that introducing \textit{both} VLL singlets and doublets and enforcing them to mix is necessary to obtain scenarios providing an explanation for the deviation between the SM predictions for $a_\mu$ and the associated measured value. It is worth noting, however, that several of these scenarios still exhibit VLLs with masses smaller than the existing collider limits, and are thus excluded phenomenologically.

Next, we present predictions for the VLL contributions to $\Delta a_e$ in the considered scenarios in figure~\ref{fig:2}, investigating their dependence on various model parameters and properties while comparing them with experimental measurements conducted on Rb atoms~\cite{Fan:2022eto}. A noticeable contrast emerges relative to the muon case. In the top-left and top-right panels of the figure, we analyse the dependence of $\Delta a_e$ on the first-generation vector-like doublet mass parameter $M^{11}_{\mathcal{D}_1}$ and on the lepton-VLL doublet Yukawa coupling $|\mathbf{\hat{Y}_{\mathcal{D}_1}}^{11}|$, respectively. We observe that scenarios in which the new physics contributions to $a_e$ align with data predictions exist within the entire scanned range for these parameters. Furthermore, as indicated by the colour coding, many of these solutions also offer a potential explanation for the observed discrepancies in the anomalous magnetic moment of the muon, as expected given the different parametric dependence of $a_\mu$ and $a_e$. In the bottom row of figure~\ref{fig:2}, we further demonstrate the existence of solutions spanning a range of masses of electrically charged and neutral VLLs coupled to first-generation ordinary leptons, with a significant portion being excluded by collider bounds. Furthermore, acceptable scenarios feature diverse VLL composition, as indicated by the associated $|\mathbf{\hat{Y}_{\mathcal{D}_1}}^{11}|/|\mathbf{\hat{Y}_{\mathcal{S}_1}}^{11}|$ ratios that can be either large or small. However, the doublet Yukawa coupling is again found to be always smaller than the singlet one.

\begin{figure*}
  \centering 
  \includegraphics[width=\textwidth]{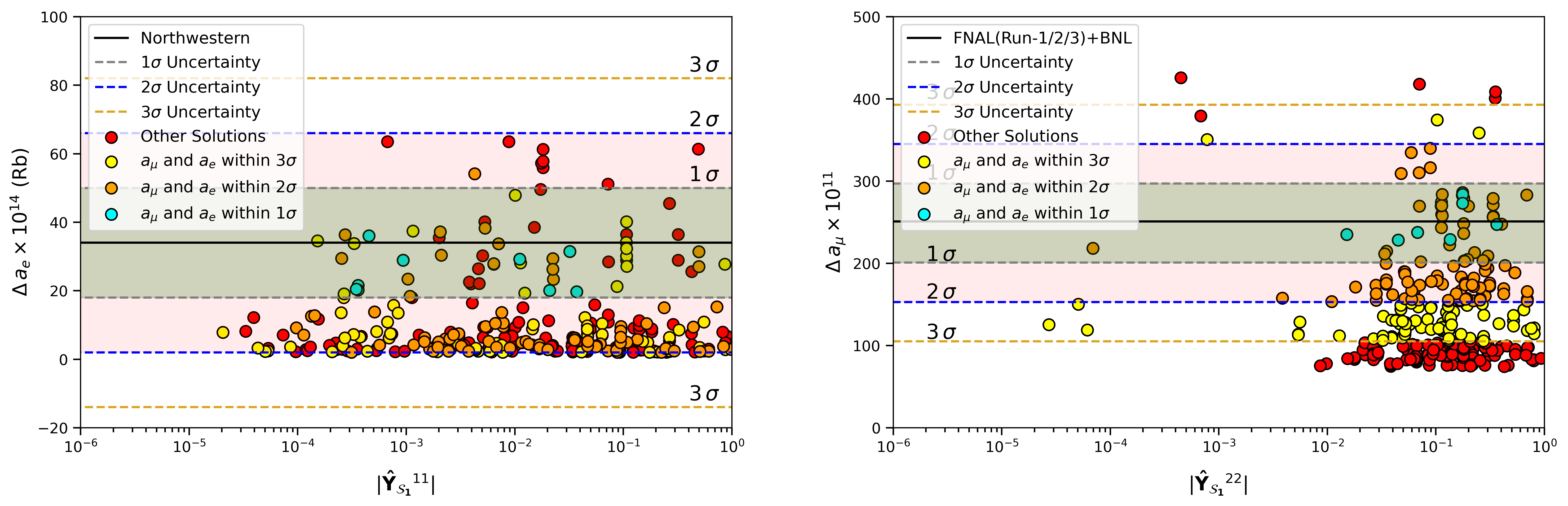}
  \caption{Predictions for the new physics contributions to the anomalous magnetic moment of the electron (left) and the muon (right) and their dependence on the relevant couplings ($|\mathbf{\hat{Y}_{\mathcal{S}_1}}^{11}|$ and $|\mathbf{\hat{Y}_{\mathcal{S}_1}}^{22}|$ respectively) of the additional VLL singlet with the ordinary SM charged lepton singlet. The colour coding depicts how likely the selected scenarios can provide a solution to both the $a_e$ and $a_\mu$ puzzles.\label{fig:3}}
\end{figure*}

To complement our exploration of the significance of incorporating charged VLL singlets $\mathcal{S}_1$ in the model, figure~\ref{fig:3} illustrates the dependence of the new physics contributions to $a_e$ (left panel) and $a_\mu$ (right panel) and their correlation with the magnitude of the deviation relative to measurements conducted at Northwestern on Rb atoms and the combination of FNAL and BNL data, respectively. Our analysis reaffirms previous conclusions, indicating that the Yukawa coupling linking a vector-like singlet of a given generation to the corresponding SM lepton doublet must be large in the case of the muon, with $|\mathbf{\hat{Y}_{\mathcal{S}_1}}^{22}| > 0.01$. However, for $a_e$, $|\mathbf{\hat{Y}_{\mathcal{S}_1}}^{11}|$ can span the entire considered set of coupling values, ranging from $10^{-5}$ to $1$. Furthermore, our findings demonstrate the feasibility of constructing scenarios that achieve agreement between theory and data for both the electron and the muon anomalous magnetic moments.

\begin{figure*}
  \centering 
  \includegraphics[width=.49\textwidth]{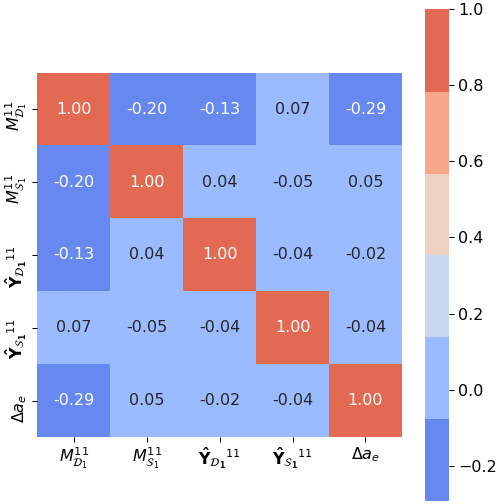} 
  \includegraphics[width=.49\textwidth]{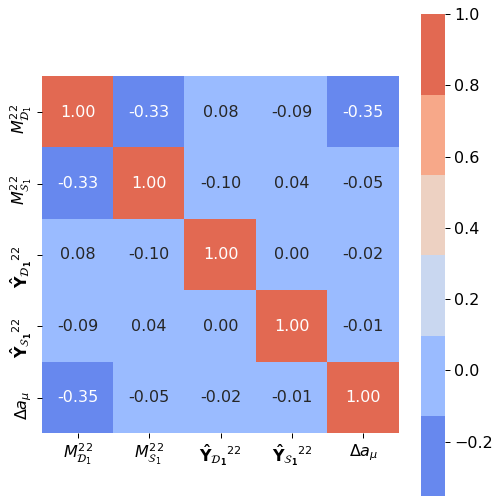}
  \caption{Heat map based on the pairwise Pearson correlation coefficients between the quantities
  $\Delta a_e$, $M^{11}_{\mathcal{D}_1}$, $M^{11}_{\mathcal{S}_1}$, $\mathbf{\hat{Y}_{\mathcal{D}_1}}^{11}$ and $\mathbf{\hat{Y}_{\mathcal{S}_1}}^{11}$ (left), as well as that computed from $\Delta a_\mu$, $M^{22}_{\mathcal{D}_1}$, $M^{22}_{\mathcal{S}_1}$, $\mathbf{\hat{Y}_{\mathcal{D}_1}}^{22}$ and $\mathbf{\hat{Y}_{\mathcal{S}_1}}^{22}$ (right).\label{fig:4}}
\end{figure*}

To further explore the impact of all model's free parameters on $\Delta a_{\mu}$ and $\Delta a_e$, we now examine how they correlate with predictions for the value of the electron and muon anomalous magnetic moments. Figure~\ref{fig:4} shows heat maps based on the pairwise Pearson correlation coefficients~\cite{JamesISL, Lista:2017jsy} between these parameters, and $\Delta a_e$ (left) and  $\Delta a_\mu$ (right). Each cell in these heat maps includes the corresponding correlation values (between $-1$ and $1$), and is colored accordingly.   The correlations are obtained from a sample containing the ensemble of the 328 scenarios identified by our scanning procedure, each scenario contributing as one input vector that includes the value of each independent parameter of the model (\textit{e.g.}, $\mathbf{\hat{Y}}_{\nu}$, $\mathbf{\hat{Y}}_{\mathcal{S}_1}$, \textit{etc}.) along with those of the observables $\Delta a_{\mu}$ and $\Delta a_{e}$ calculated at one loop by {\tt SPheno}. In practice, Pearson correlation coefficients are computed with a homemade {\tt Python} program using a specific {\tt NumPy} \cite{harris2020array} array of 328 elements for each component of the input vector, the data being collected within a {\tt pandas} \cite{reback2020pandas,mckinney-proc-scipy-2010} data frame. Then we use {\tt StandardScalar} class within the {\tt preprocessing} package of the {\tt sklearn} library~\cite{scikit-learn} to standardize whole the dataset, implying that large values of the input parameters can be compared to small values within a correlation calculation. Employing the standardized dataset, we finally proceed with calculating the correlations by applying the {\tt corr} method to the collected data frame, and generate the plots with the {\tt heatmap} function of the {\tt seaborn} library \cite{Waskom2021}.

To elaborate on the employed correlation formula, we remind that Pearson correlation coefficients measure the linear relationship between two quantities by calculating the ratio of their covariance to the product of their standard deviations. According to this definition, the coefficient lies within the range $[-1,1]$, where a value equal to $1$ suggests a perfect positive linear relationship (the two quantities exhibiting a similar behaviour), and $-1$ a perfect negative linear relationship (the two quantities exhibiting an opposite behaviour), while $0$ represents no linear relationship. Thus the heat maps in figure~\ref{fig:4} indicate the linear relationships of $\Delta a_\mu$ and $\Delta a_e$ with the mass and Yukawa terms of the Lagrangian. However, such coefficients ignore other types of (non-linear) relationships between quantities. The corresponding formula for calculating the Pearson coefficients between two parameters $x$ and $y$ across a given dataset is given by
\begin{align}\label{electron-g-2-Cs}
   \rho\,(x,y)\,=\,\frac{\text{cov}(x,y)}{\sigma_x\,\sigma_y}\,,
\end{align}
where $\mathrm{cov}(x,y)$ denotes the covariance between the two parameters across the dataset, and $\sigma_x$ and $\sigma_y$ the standard deviation obtained for the parameters $x$ and $y$ respectively.

The results depicted in figure~\ref{fig:4} highlight the significance of the parameter $M^{11}_{\mathcal{D}_1}$ in determining the VLL contributions to the anomalous magnetic moment of the electron. Across all 328 solutions identified in our scan, we observe a Pearson correlation of $-0.29$ between $M^{11}_{\mathcal{D}_1}$ and $\Delta a_e$. This negative correlation suggests that increasing the mass parameter associated with the first-generation VLL doublet leads to a decrease in the new physics contribution to the electron's anomalous magnetic moment. To reconcile this with experimental data, a spectrum featuring lighter VLL mass eigenstates may be required, as other coupling parameters play no role. This is achievable through lepton-VLL mixings. Conversely, regions within the parameter space characterised by lower $M^{11}_{\mathcal{D}_1}$ values are favoured in light of considerations regarding $a_e$, which is consistent with findings from prior studies (as \textit{e.g.}\ in~\cite{Dermisek:2013gta}). In the right panel of figure~\ref{fig:4}, we demonstrate that a similar correlation exists for $\Delta a_{\mu}$ and the second-generation parameter $M^{22}_{\mathcal{D}_1}$. 

A notable feature from figure~\ref{fig:4} is the absence of a clear hierarchy between the Yukawa interactions linking VLL doublets and singlets to the SM ordinary leptons when predictions for the anomalous magnetic moment are enforced to match experimental data. The Yukawa couplings $\mathbf{\hat{Y}_{\mathcal{S}_1}}^{11}$ and $\mathbf{\hat{Y}_{\mathcal{D}_1}}^{11}$ ($\mathbf{\hat{Y}_{\mathcal{S}_1}}^{22}$ and $\mathbf{\hat{Y}_{\mathcal{D}_1}}^{22}$) exhibit no discernible linear correlation with $a_e$ ($a_\mu$), suggesting that any hierarchy among them may be suitable.  This can also be seen from figures~\ref{fig:1}, \ref{fig:2} and \ref{fig:3} where we note that the Yukawa couplings of the VLL singlets to the SM leptons consistently exceed those of the VLL doublets.

%%%%%%%%%%%%%%%%%%%%%%%%%%%%%%%%%%%%%%%%%%%
\subsection{Properties of the neutrino sector}
\label{subsec:neutrinos}
%%%%%%%%%%%%%%%%%%%%%%%%%%%%%%%%%%%%%%%%%%%%%

The introduction of VLLs has implications for the neutrino sector, as it can impact the mechanism to generating neutrino masses. We must therefore ensure that the model is compatible with neutrino data. While compelling evidence from oscillation experiments indicates that their mass is non-zero~\cite{SNO:2002tuh, Super-Kamiokande:2005mbp}, the actual mass spectrum is unresolved. It can nevertheless be constrained thanks to cosmology. Indeed, the abundance of cosmic neutrinos and their contribution to the universe's total energy density during its early stages enables the study of their characteristics. Specifically, the sum of the neutrino masses $\sum m_{\nu}$ influences the cosmic microwave background, supernov\ae\ features, large scale structure, and big bang nucleosynthesis~\cite{Loureiro:2018pdz}. Present data has yielded an approximate upper limit of $\sum m_{\nu} < 0.26$~eV, a constraint that we have incorporated in our parameter space scans (see section~\ref{sec:model}). In our analysis of the neutrino sector, we have considered scenarios with the normal ordering (NO) and inverted ordering (IO) of the Majorana neutrino masses, and we have explored all mixing possibilities of the SM neutrinos with the neutral components of the VLL doublets and neutral singlets. The constraints imposed on the neutrino sector during the scan are detailed in Table~\ref{tab:3}.

\begin{table}
  \begin{center}\renewcommand{\arraystretch}{1.3}\setlength{\tabcolsep}{9pt}
  \begin{tabular}{c}
	Cosmological constraints\\
	$\Sigma\,m_{\nu}\,\leq \,$0.26 eV\\[.2cm]\hline 
 	Neutrino masses (IO)\\
	$m_{\nu_1}\in[8.37,8.85]\times 10^{-3}$ eV\\ 
	$m_{\nu_2}\in[4.93,5.04]\times 10^{-2}$ eV\\[.2cm]\hline 
    Neutrino masses (NO)\\
	  $m_{\nu_2}\in[8.46,8.88]\times 10^{-3}$ eV \\ 
	$m_{\nu_3}\in[4.88,5.02]\times 10^{-2}$ eV  
  \end{tabular}
  \caption{Constraints imposed on the neutrino sector during our parameter space scan. We have allowed a $2\sigma$ uncertainty on the neutrino masses, based on the experimental values determined for the neutrino squared mass differences $\Delta m_{12}^2$ and $\Delta m_{31}^2$.\label{tab:3}}
  \end{center}
\end{table}

\begin{figure*}
\centering 
  \includegraphics[width=\textwidth]{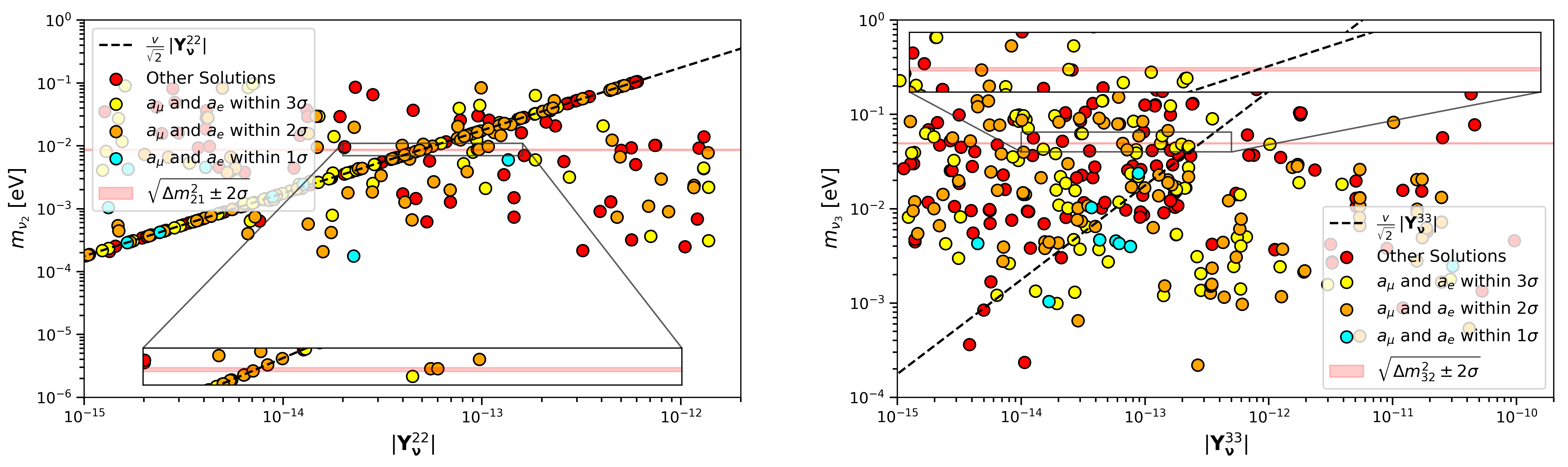}\\
  \includegraphics[width=\textwidth]{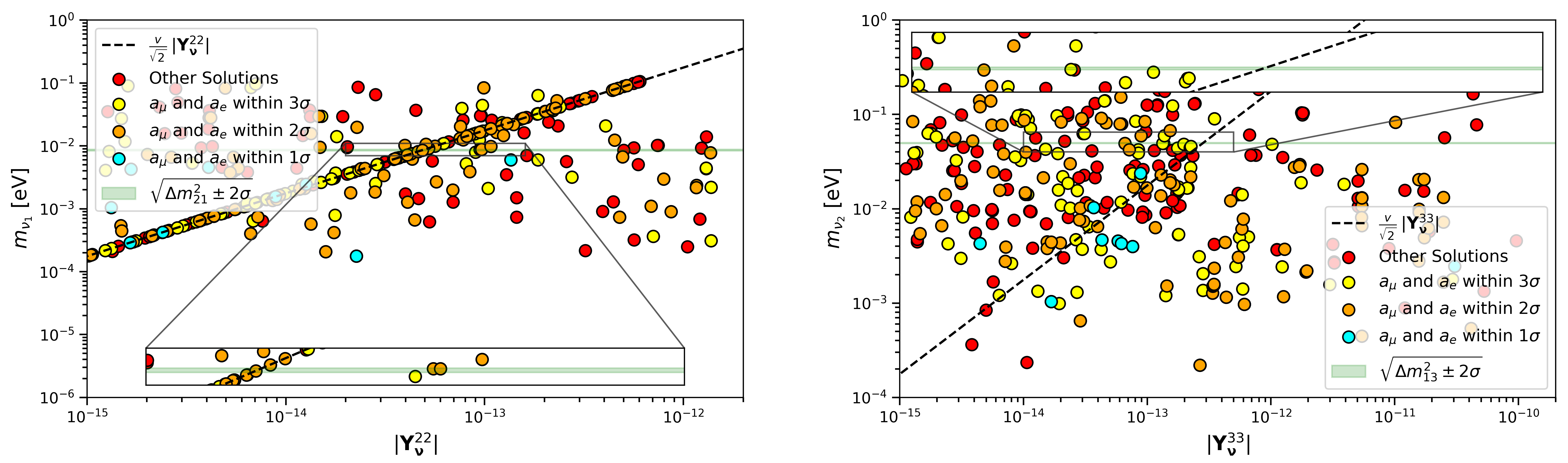}
  \caption{Predictions for the masses of the heavier neutrinos at one loop, for scenarios featuring a normal ordering of the neutrino masses (top row) and an inverted ordering of the neutrino masses (bottom row). The results are shown as a function of the neutrino Yukawa couplings $\mathbf{Y}_\mathbf{\nu}^{22}$ (left) and $\mathbf{Y}_\mathbf{\nu}^{33}$ (right), while the dotted lines correspond to tree-level predictions. The narrow red and green bands depict the range of masses consistent with neutrino oscillation data within $2\sigma$. \label{fig:6}}
\end{figure*}

\begin{figure*}
\centering
  \includegraphics[width=.49\textwidth,valign=c]{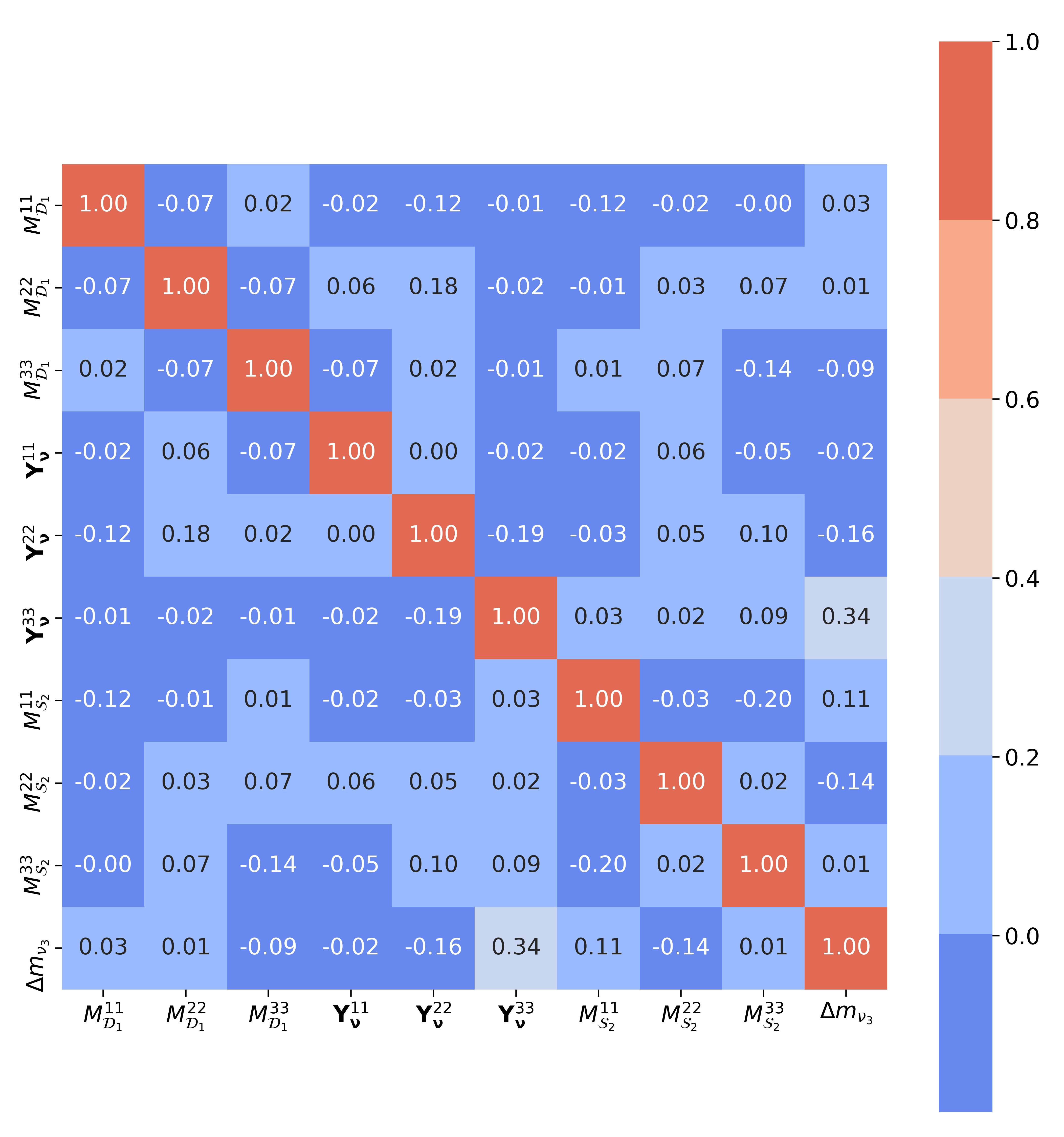}
  \includegraphics[width=.50\textwidth,valign=c]{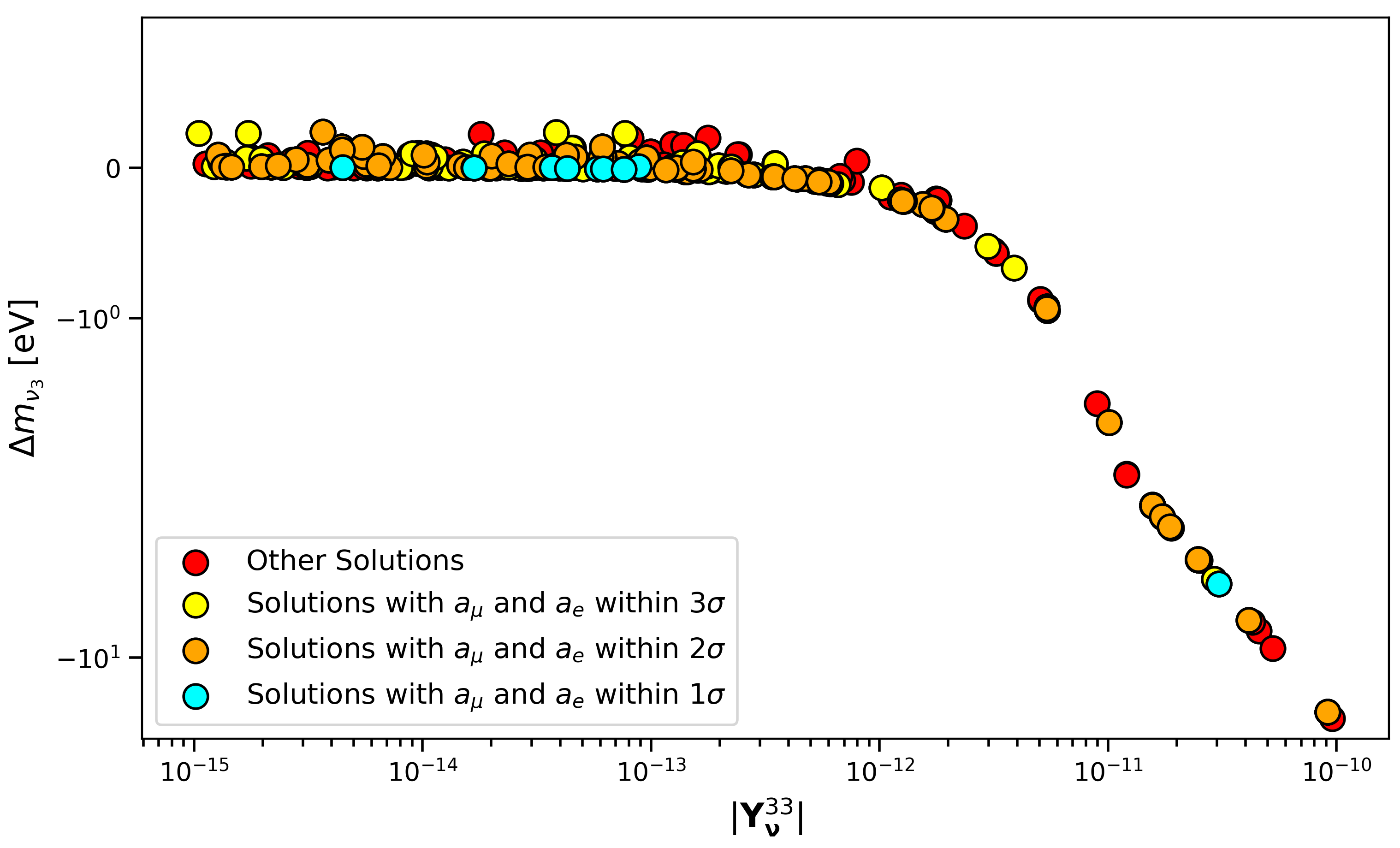}
  \caption{Heat map based on the pairwise Pearson correlation coefficients between the mass of the third neutrino $m_{\nu_3}$ and the parameters entering its calculation at one loop (left). The effects of the most impactful parameter $\mathbf{Y}_\mathbf{\nu}^{33}$ on the corrections to $m_{\nu_3}$, denoted by $\Delta m_{\nu_3}$, is additionally shown (right).\label{fig:7}}
\end{figure*}

In figure \ref{fig:6}, we analyse the results of our parameter space scan in relation to their impact on neutrino masses. We classify the solutions into those with normal (top row) and inverted (bottom row) neutrino mass ordering, and we display the dependence of the masses of the heavier neutrinos on the neutrino Yukawa couplings to ordinary leptons $\mathbf{Y}_\mathbf{\nu}^{22}$ (left panels) and $\mathbf{Y}_\mathbf{\nu}^{33}$ (right panels). We recall that for normal-ordering scenarios, the lightest neutrino is $\nu_1$, while for inverted-ordering scenarios, it is $\nu_3$. Additionally, we use colour coding to highlight how each solution aligns with observed deviations in the anomalous magnetic moments of the muon and the electron. Many solutions exhibit a direct proportionality to the Yukawa coupling $\mathbf{Y}_\mathbf{\nu}$, as indicated by the dashed lines shown in the different figures. In such cases, loop corrections to neutrino masses, that involve the VLL states of the model, are negligibly small. However, this is not systematic, and when loop corrections are not negligible, results show a slight dependence on the magnitude of the neutrino Yukawa couplings. We also represent solutions compatible with the constraints on the neutrino masses listed in Table~\ref{tab:3} by using narrow green and red bands. Our analysis reveals that solutions explaining observed deviations observed in $a_\mu$ and $a_e$, as well as generating neutrino masses in agreement with data, exist for both normal and inverted neutrino mass orderings.

In order to assess which model parameters have the most significant impact on the neutrino sector, we present a heat map in the left panel of figure~\ref{fig:7} based on the pairwise Pearson correlation coefficients between the mass of the neutrino $\nu_3$ (chosen for illustrative purposes) and various model parameters relevant for neutrino masses at one loop. It turns out that the parameter $\mathbf{Y}_\mathbf{\nu}^{33}$, governing the Yukawa interactions between ordinary leptons and the right-handed neutrino field, has the largest effect. Turning our attention to the value of the one-loop corrections $\Delta m_{\nu_3}$ to the tree-level mass, we observe, in the right panel of the figure, that the value of $\mathbf{Y}_\mathbf{\nu}^{33}$ plays a minimal role on the neutrino mass for $\mathbf{Y}_\mathbf{\nu}^{33} \lesssim 10^{-12}$. However, for larger values, these effects become more pronounced as one-loop contributions $\Delta m_{\nu_3}$ must compensate for a too large tree-level mass. This easily leads to predictions that conflict with neutrino oscillation data, as visible from the figure. Therefore, achieving agreement with experimental constraints necessitates a very weak interaction term between the right-handed neutrino field $\nu$ and the SM-like lepton doublet $L$. Furthermore, our results demonstrate that while neutrino data can be accommodated in a satisfactory manner, the additional degrees of freedom provided by the VLL sector extend the allowed range for the neutrino Yukawa couplings. This could pave the way to a theoretically-acceptable solution to the neutrino mass problematic with larger Yukawa coupling values. The tiny values of the Yukawa couplings indeed demonstrate by themselves that right-handed neutrinos alone cannot be a reason for explaining the neutrino data, as such small values are theoretically not elegant.

\begin{figure*}
  \centering 
  \includegraphics[width=.7\textwidth]{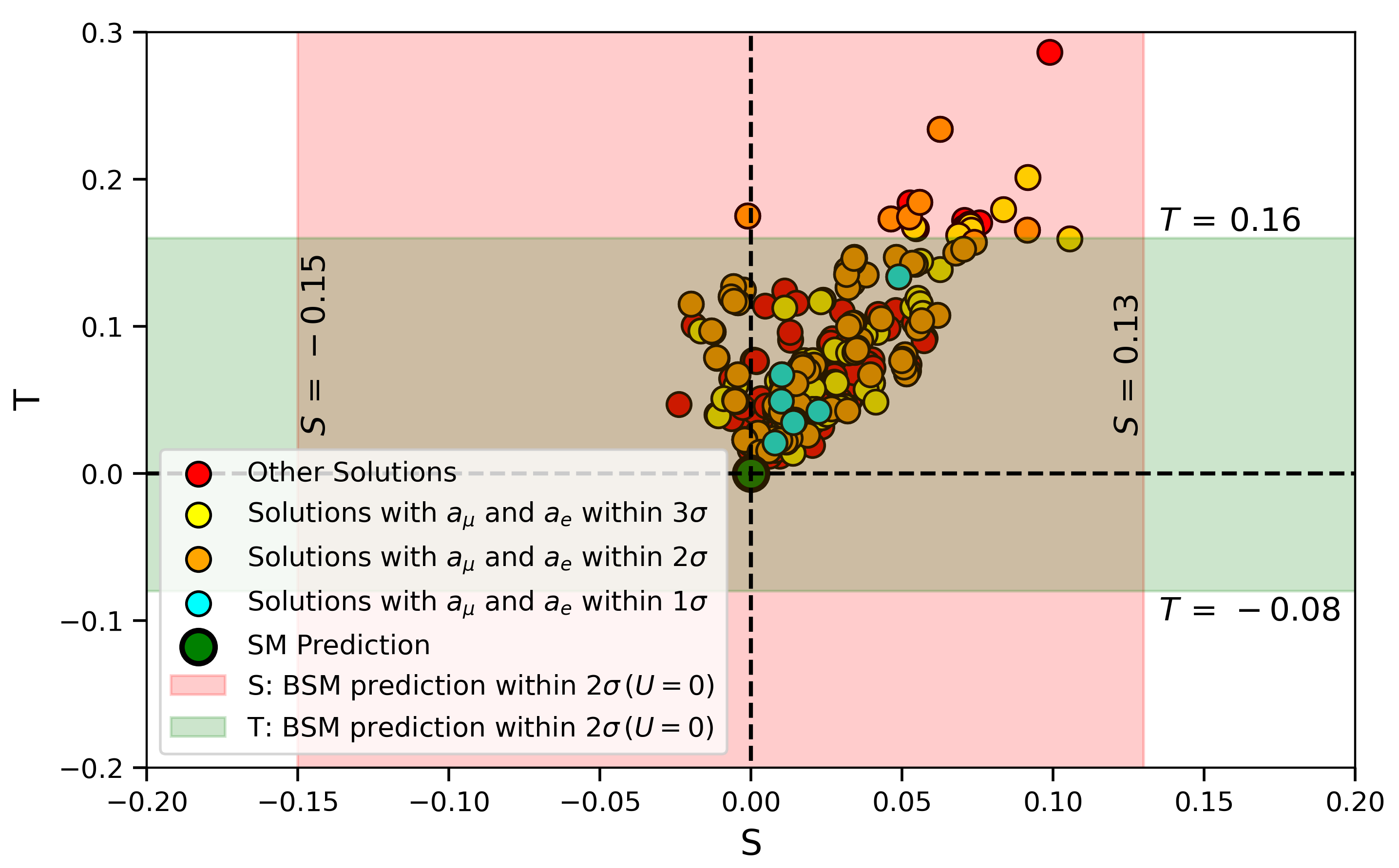}
  \caption{New physics contributions to the $S$ and $T$ parameters, and comparison with current limits at $2\sigma$, shown in red and green respectively. Colour coding indicates the consistency of the solutions chosen in the scan with respect to their potential to explain deviation in the anomalous magnetic moments of the muon and the electron. \label{fig:9}}
\end{figure*}

%%%%%%%%%%%%%%%%%%%%%%%%%%%%%%%%%
\subsection{Electroweak precision tests}
\label{subsec:oblique}
%%%%%%%%%%%%%%%%%%%%%%%%%%%%%%%%%

Given that our theoretical framework includes a non-trivial $SU(2)_L$ multiplet, we perform an additional verification to ensure that the selected scenarios in our scan comply with electroweak precision data. We undertake this verification by calculating the new physics contributions to the Peskin-Takeuchi parameters~\cite{Peskin:1991sw}. In light of recent bounds on these parameters ($S=-0.01 \pm 0.07$, $T=0.04 \pm 0.06$, and $U=0$~\cite{Workman:2022ynf} ), we show in figure~\ref{fig:9} that the majority of scenarios compatible with neutrino oscillation data and providing an explanation for the anomalous magnetic moment of the muon and the electron do not violate electroweak precision data.

%%%%%%%%%%%%%%%%%%%%%%%%%%%%%%%%%%%%%%%%%%%%%%%%%%%%%%%%%%%%%%%%
\section{Collider phenomenology}
\label{sec:collider}
%%%%%%%%%%%%%%%%%%%%%%%%%%%%%%%%%%%%%%%%%%%%%%%%%%%%%%%%%%%%%%%%
In this section, our focus is on analysing the phenomenological implications at high-energy hadronic colliders of scenarios that meet all previously imposed constraints. We examine two representative benchmark setups and demonstrate that they yield discernible signals amidst the overwhelming background of the SM~\cite{Bissmann:2020lge}. For the first scenario (see section~\ref{subsec:BM1}), we exploit a signature characterised by the production of a significant number of charged leptons, while for the second one (see section~\ref{subsec:BM2}), we leverage the production of substantial missing energy. In general, studying these two signals is sufficient to obtain excellent collider handles on model scenarios consistent with all constraints examined in section~\ref{sec:scanresults}.

To conduct our study, we generate a UFO model~\cite{Darme:2023jdn} with the {\tt SARAH} package (version 4.15.1)~\cite{Staub:2013tta, Goodsell:2017pdq}, starting from the model implementation discussed in section~\ref{sec:model}. This allows us to utilise {\tt MadGraph5\_aMC@NLO}~\cite{Alwall:2014hca} (version 3.2.0) to generate both signal and background events relevant to a potential proton-proton collider operating at a centre-of-mass energy of 100~TeV, and aiming to collect an integrated luminosity of 3~ab$^{-1}$. We obtain hard-scattering events by convoluting leading-order matrix elements with the next-to-leading-order set of NNPDF 2.3 parton densities~\cite{Ball:2012cx}. Subsequently, these events are matched with parton showering using {\tt Pythia} 8.3~\cite{Bierlich:2022pfr}, which we also use to simulate hadronisation. Our simulations match matrix elements including up to one additional parton at the Born level, using the MLM matching scheme~\cite{Mangano:2006rw, Alwall:2008qv}, and we re-weight events to incorporate higher-order corrections in QCD into the relevant total rates~\cite{Alwall:2014hca, Gauld:2021ule, AH:2023hft}. Finally, hadron-level events are processed through the fast detector simulator {\tt SFS}~\cite{Araz:2020lnp} equipping the {\tt MadAnalysis}~5 package~\cite{Conte:2012fm, Conte:2014zja, Conte:2018vmg} (version 1.10.12). This internally relies on {\sc FastJet}~\cite{Cacciari:2011ma} (version 3.3.3) for event reconstruction, together with its implementation of the anti-$k_T$ jet algorithm~\cite{Cacciari:2008gp} with a radius parameter $R = 0.4$. To emulate realistic detector resolution and particle identification, we employ the ATLAS configuration card provided with the {\tt SFS} module of {\tt MadAnalysis}~5, which we also use for implementing our collider analysis.

%%%%%%%%%%%%%%%%%%%%%%%%%%%%%%%%%
\subsection{A signature involving multiple charged leptons}\label{subsec:BM1}
%%%%%%%%%%%%%%%%%%%%%%%%%%%%%%%%%

\begin{table*}
  \begin{center}\renewcommand{\arraystretch}{1.3}
  \begin{tabular}{cc@{\hspace{10pt}}|@{\hspace{10pt}}cc@{\hspace{10pt}}|@{\hspace{10pt}}cc}
    $\mathbf{\hat{Y}_{\mathcal{D}_1}}^{11}$ & $1.34\times 10^{-1}$  & $\mathbf{\hat{Y}_{\mathcal{S}_1}}^{11}$ & $-1.15\times 10^{-3}$ & $M^{11}_{\mathcal{S}_2}$ & $400.47$ GeV\\
    $\mathbf{\hat{Y}_{\mathcal{D}_1}}^{22}$ & $-3.18\times 10^{-2}$  & $\mathbf{\hat{Y}_{\mathcal{S}_1}}^{22}$ & $6.43\times 10^{-1}$ & $M^{22}_{\mathcal{S}_2}$ & $435.47$ GeV\\
    $\mathbf{\hat{Y}_{\mathcal{D}_1}}^{33}$ & $-3.06\times 10^{-1}$  & $\mathbf{\hat{Y}_{\mathcal{S}_1}}^{33}$ & $-1.75\times 10^{-5}$ & $M^{33}_{\mathcal{S}_2}$ & $156.56$ GeV\\ 
    $(\lambda_{L}^{(1)})^{11}$ & $-1.38\times 10^{-4}$ & $(\lambda_{R}^{(1)})^{11}$ & $-8.50\times 10^{-2}$ & $M^{11}_{\mathcal{D}_1}$ & $131.39$ GeV \\ 
    $(\lambda_{L}^{(1)})^{22}$ & $-2.86\times 10^{-1}$ & $(\lambda_{R}^{(1)})^{22}$ & $6.55\times 10^{-4}$ & $M^{22}_{\mathcal{D}_1}$ &  $125.97$ GeV \\ 
    $(\lambda_{L}^{(1)})^{33}$ & $-3.29\times 10^{-5}$ & $(\lambda_{R}^{(1)})^{33}$ & $-5.29\times 10^{-1}$ & $M^{33}_{\mathcal{D}_1}$ & $178.34$ GeV \\
    $\mathbf{\hat{Y}_{\mathcal{S}_2}}^{11}$ & $-7.62\times 10^{-5}$ & $\mathbf{Y}_\mathbf{\nu}^{11}$ & $-2.81\times 10^{-13}$ & $M^{11}_{\mathcal{S}_1}$&  $391.12$ GeV \\
    $\mathbf{\hat{Y}_{\mathcal{S}_2}}^{22}$ &  $9.62\times 10^{-7}$ & $\mathbf{Y}_\mathbf{\nu}^{22}$ & $3.53\times 10^{-15}$ & $M^{22}_{\mathcal{S}_1}$&  $431.56$ GeV\\
    $\mathbf{\hat{Y}_{\mathcal{S}_3}}^{11}$ & $-1.17\times 10^{-7}$  & $\mathbf{Y}_\mathbf{\nu}^{33}$ & $1.33\times 10^{-13}$ & $M^{33}_{\mathcal{S}_1}$& $177.65$ GeV\\
    $(\lambda_{L,R}^{(2)})^{11}$ & $2.15\times 10^{-7}$ &
    $(\lambda_{L,R}^{(2)})^{22}$ & $-1.93\times 10^{-5}$ &
    $(\lambda_{L,R}^{(2)})^{33}$ & $-4.10\times 10^{-7}$ \\ \hline
    $M_h$ & 127.66 GeV & $m_{e^-_4}$ & 126.12 GeV & $m_{e^-_5}$ & 134.23 GeV \\
    $m_{e^-_6}$ & 144.01 GeV & $m_{e^-_9}$ & 448.21 GeV & $m_{\nu_1}$ & $6.18 \times 10^{-4}$ eV\\
    $m_{\nu_2}$ & $4.51 \times 10^{-3}$ eV & $m_{\nu_3}$ & $4.94 \times 10^{-2}$ eV& $S$ & $2.33 \times 10^{-2}$ \\
    $T$ & $3.81 \times 10^{-2}$ & $\Delta a_e$ & $3.74 \times 10^{-13}$& $\Delta a_{\mu}$ & $1.44 \times 10^{-9}$ \\
  \end{tabular}
  \caption{Values of the free parameters of the model for the benchmark scenario {\bf BM I}, together with predictions for the most relevant observables used in our parameter space exploration.\label{tab:4}}
  \end{center}
\end{table*}

As a first representative benchmark scenario ({\bf BM I}) satisfying all constraints imposed, we consider a choice of parameters leading to a new physics spectrum featuring heavy VLLs with masses ranging in about $[125, 450]$~GeV with a collider signature leading to enriched muon production. The corresponding model parameters are given in Table \ref{tab:4}, together with the numerical values for the observables relevant for the constraints that we imposed during the scan. In particular, we observe that predictions for the oblique parameters are compliant with the latest observations, that the sum of the masses of the three lightest neutrinos, for normal ordering, satisfies cosmological bounds, and that the {\bf BM I} scenario provides an explanation for observed deviations in the anomalous magnetic moments of both the electron and the muon at $2\sigma$ and $1\sigma$ respectively. While many of the six electrically-charged VLL eigenstates of the spectrum have masses conducive to significant production rates at colliders, only a select few leave discernible signatures once decay rates are accounted for. Among these, the heaviest VLL, denoted as $e^-_9$ (as the heaviest of all nine charged leptons, with $e_4^-$ being, in this notation, the lightest of all electrically-charged BSM leptonic states), emerges as the leading candidate.

Among the various production and decay processes involving VLLs, the most promising collider signature involves multiple leptons. This signature stems from the electroweak production of a pair of VLLs, each decaying into a Standard Model lepton and a weak boson. We choose to focus on the leptonic decay channel of the weak bosons to maximise the number of leptons in the final state, since final states with four or more leptons typically exhibit minimal or negligible SM background. By considering predictions for the VLL electroweak couplings, masses, and branching ratios within the framework of benchmark scenario \textbf{BM I}, we identify the production of a pair of $e_9$ states decaying into a muon and a $Z$-boson as the most promising channel,
\begin{equation}\bsp
  p\,\, p \ \ \rightarrow& \ \  e^-_9\,\,e^+_9 + X \ \ \rightarrow\ \   \mu^-\,Z \, \mu^+\,\,Z + X \\
   &\quad \ \ \rightarrow \mu^- \ell^+\ell^-\, \mu^- \ell^{\prime+}\ell^{\prime-} + X\,.
\esp\label{BMI_process}\end{equation}
This process is associated with a production rate at next-to-leading order (NLO) and includes soft-gluon resummation at the next-to-next-to-leading-logarithmic (NNLL) accuracy of 26.4~fb~\cite{AH:2023hft}, and with a VLL branching ratio $\mathrm{BR}(e_9 \to \mu\, Z) = 47.7\,$\%. This indicates a significant probability of producing events with six isolated leptons after accounting for leptonic $Z$-boson decays, which is largely free from SM background. Additionally, the presence of two hard muons originating from the decay of $e_9$ VLLs with masses in the several hundred GeV range provides an additional discriminating feature for the signal. In the subsequent analysis, we provide quantitative evidence that investigating such a process is sufficient for a potential observation of a VLL signal at a future proton-proton collider operating at a centre-of-mass energy $\sqrt{s}=100 \, \text{TeV}$, when the properties of this VLL state are such that they provide an explanation for several leptonic anomalies. For this, we perform signal simulation as described above, with a merging scale set to $m_{e_9}/4$.

\begin{figure*}
  \centering 
  \includegraphics[width=.49\textwidth, trim={1cm 13.7cm 19.5cm 12.2cm},clip]{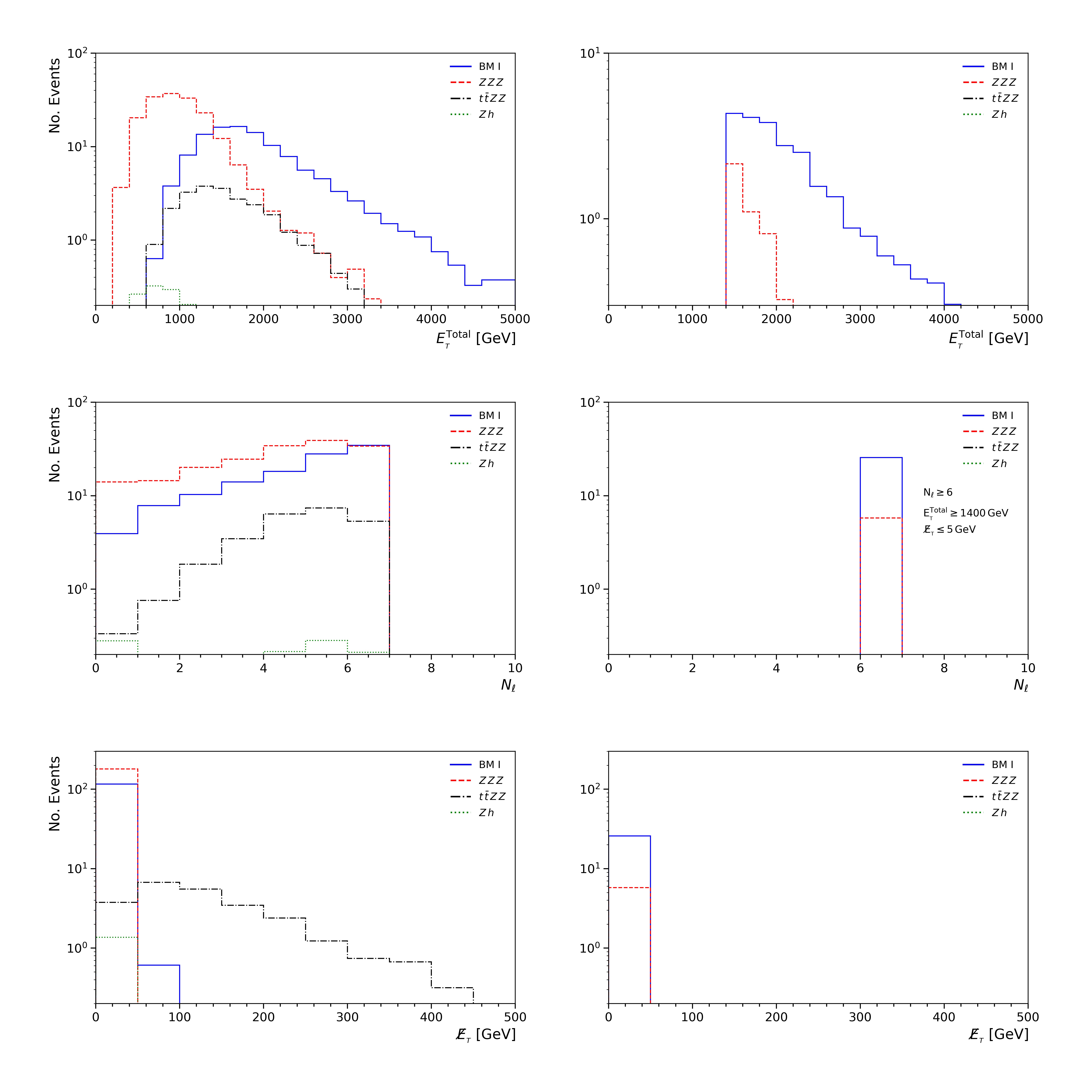}  
  \hfill
  \includegraphics[width=.49\textwidth, trim={1cm 25.9cm 19.5cm 0},clip]{figures/ppe9e9barBMI}
  \caption{Key observables for distinguishing the \textbf{BM I} signal from the SM background. We display the distribution of the number of reconstructed leptons $N_\ell$ (left), and of the total transverse energy $E_T^\mathrm{Total}$ (right). The signal is plotted in solid lines, the background in dashed lines. Histograms are normalised to a luminosity of $3\,\text{ab}^{-1}$ of $pp$ collisions at $\sqrt{s} = 100\,\text{TeV}$.} \label{fig:10}
\end{figure*}

While we anticipate a small SM background, we nonetheless perform its detailed phenomenological analysis, including event generation for all relevant SM processes using the previously introduced toolchain. These processes consist of:
\begin{equation}\bsp
  &p\,\, p \ \rightarrow\  h\,\,Z +X \,,\quad\\
  &p\,\, p \ \rightarrow\  Z\,\,Z\,\,Z +X\,,\quad\\
  &p\,\, p \ \rightarrow\ t\,\,\bar{t}\,\,Z\,\,Z + X\,,
\esp\end{equation}
which all lead to a six-lepton final state once $Z$-boson and Higgs-boson leptonic decays are accounted for. The generated events are re-weighted to ensure that the corresponding hard-scattering cross sections, including the branching ratios related to leptonic Higgs and $Z$ boson decays, are accurate to next-to-next-to-leading order (NNLO) for the first process~\cite{Gauld:2021ule}, and NLO for the last two processes~\cite{Alwall:2014hca}. This yields cross section values of $\sigma_{Zh} = 0.455~\text{ab}$, $\sigma_{ZZZ} = 0.0602~\text{fb}$ and $\sigma_{t\bar{t}ZZ} = 0.371~\text{ab}$, that we determine by multiplying the leading-order predictions returned by \texttt{MadGraph5\_aMC@NLO} by constant $K$-factors extracted from the literature.

In our analysis, we begin by defining a lepton collection comprising all electrons and muons well-separated from any jet candidate by a transverse distance $\Delta R>0.4$, with a pseudo-rapidity $ |\eta^\ell| \leq 2.5$, and a transverse momentum $p_T^\ell > 15$~GeV. Additionally, we consider as jet candidates all reconstructed jets with a pseudo-rapidity $ |\eta^j| \leq 2.5$ and a transverse momentum $p_T^j > 15$~GeV. We then select events containing $N_\ell \geq 6$ reconstructed leptons and a total transverse energy $E_T^\mathrm{Total}$ defined as the sum of the $p_T$ of all reconstructed objects of at least 1.4~TeV. Figure~\ref{fig:10} illustrates the potential of these criteria for a high signal efficiency while effectively rejecting background events. After applying these criteria, we select $n_s$ signal events and $n_b$ background events, where $n_b$ is determined by the sum of the number of surviving $Zh$, $ZZZ$ and $t\bar{t}ZZ$ events ($n_{Zh}$, $n_{ZZZ}$ and $n_{t\bar{t}ZZ}$, respectively). Specifically, we find
\begin{equation}
  n_s = 28.43
  \qquad\text{and}\qquad
  n_b = 9.24
\end{equation}
with
\begin{equation}
  n_{Zh} \simeq 0,\ \
  n_{ZZZ} = 6.23 \ \ \text{and}\ \
  n_{t\bar{t}ZZ} = 2.99\,.
\end{equation}
We estimate the sensitivity to the signal using two metrics~\cite{Cowan:2010js},
\begin{equation}
s  = \frac{n_s}{\sqrt{n_b+\Delta_b^2}}\,,
\end{equation}
and
\begin{equation}\bsp
Z_A =&\ \Bigg[ 2\bigg\{
 (n_s+n_b)\ln\frac{(n_s+n_b)(n_s+\Delta_b^2)}{n_b^2+(n_s+n_b)\Delta_b^2} \\
 & \hspace{2.cm} -
 \frac{n_b^2}{\Delta_b^2}\ln\bigg(1+\frac{\Delta_b^2 n_s}{n_b(n_b+\Delta_b^2)}\bigg)
 \bigg\}\Bigg]^{1/2} \ ,
\esp\end{equation}
where the systematics on the background are conservatively assumed to be $\Delta_b = 0.2 n_b$. We obtain
\begin{equation}
  s\,=\,8\,\sigma\qquad\text{and}\qquad Z_A\,=\,10\,\sigma\,,
\end{equation}
indicating a clear possibility of observing the signal at a future collider aiming to operate at a centre-of-mass energy of 100~TeV.

%%%%%%%%%%%%%%%%%%%%%%%%%%%%%%%%%
\subsection{A signature with significant missing transverse energy}\label{subsec:BM2}
%%%%%%%%%%%%%%%%%%%%%%%%%%%%%%%%%

\begin{table*}
  \begin{center}\renewcommand{\arraystretch}{1.3}
  \begin{tabular}{cc@{\hspace{10pt}}|@{\hspace{10pt}}cc@{\hspace{10pt}}|@{\hspace{10pt}}cc}
		$\mathbf{\hat{Y}_{\mathcal{D}_1}}^{11}$ & $-3.87\times 10^{-2}$  & $\mathbf{\hat{Y}_{\mathcal{S}_1}}^{11}$ & $3.13\times 10^{-4}$ & $M^{11}_{\mathcal{S}_2}$ & $152.84$ GeV\\
		$\mathbf{\hat{Y}_{\mathcal{D}_1}}^{22}$ & $-2.27\times 10^{-2}$ & $\mathbf{\hat{Y}_{\mathcal{S}_1}}^{22}$ & $-4.33\times 10^{-1}$ & $M^{22}_{\mathcal{S}_2}$ & $231.35$ GeV\\
		$\mathbf{\hat{Y}_{\mathcal{D}_1}}^{33}$ & $2.14\times 10^{-4}$ & $\mathbf{\hat{Y}_{\mathcal{S}_1}}^{33}$ & $-5.60\times 10^{-3}$ & $M^{33}_{\mathcal{S}_2}$ & $141.43$ GeV\\ 
		$(\lambda_{L}^{(1)})^{11}$ & $-1.81\times 10^{-1}$ & $(\lambda_{R}^{(1)})^{11}$ & $-2.39\times 10^{-5}$ & $M^{11}_{\mathcal{D}_1}$ & $137.82$ GeV \\ 
		$(\lambda_{L}^{(1)})^{22}$ & $9.32\times 10^{-1}$ & $(\lambda_{R}^{(1)})^{22}$ & $-9.40\times 10^{-3}$ & $M^{22}_{\mathcal{D}_1}$ &  $125.82$ GeV \\ 
		$(\lambda_{L}^{(1)})^{33}$ & $-1.76\times 10^{-3}$ & $(\lambda_{R}^{(1)})^{33}$ & $-6.34\times 10^{-2}$ & $M^{33}_{\mathcal{D}_1}$ & $296.30$ GeV \\
		$\mathbf{\hat{Y}_{\mathcal{S}_2}}^{11}$ & $-1.46\times 10^{-4}$ & $\mathbf{Y}_\mathbf{\nu}^{11}$ & $-2.79\times 10^{-13}$ & $M^{11}_{\mathcal{S}_1}$ & $622.87$ GeV \\
		$\mathbf{\hat{Y}_{\mathcal{S}_2}}^{22}$ &  $-6.28\times 10^{-5}$ & $\mathbf{Y}_\mathbf{\nu}^{22}$ & $1.01\times 10^{-15}$ & $M^{22}_{\mathcal{S}_1}$& $883.75$ GeV\\
		$\mathbf{\hat{Y}_{\mathcal{S}_2}}^{33}$ & $-3.17\times 10^{-4}$  & $\mathbf{Y}_\mathbf{\nu}^{33}$ & $6.60\times 10^{-14}$ & $M^{33}_{\mathcal{S}_1}$& $892.46$ GeV\\
		$(\lambda_{L,R}^{(2)})^{11}$ & $-5.87\times 10^{-5}$ & $(\lambda_{L,R}^{(2)})^{22}$ & $4.53\times 10^{-4}$ &  $(\lambda_{L,R}^{(2)})^{33}$ & $-1.30\times 10^{-5}$ \\ \hline
            $M_h$ & 122.57 GeV & $m_{e^-_4}$ & 110.94 GeV & $m_{e^-_5}$ & 118.70 GeV \\
            $m_{e^-_6}$ & 148.66 GeV & $m_{e^-_7}$ & 189.18 GeV & $m_{\nu_1}$ & $1.78 \times 10^{-4}$ eV\\
            $m_{\nu_2}$ & $4.92 \times 10^{-3}$ eV & $m_{\nu_3}$ & $4.86 \times 10^{-2}$ eV& $S$ & $1.02 \times 10^{-2}$ \\
            $T$ & $6.38 \times 10^{-2}$ & $\Delta a_e$ & $2.06 \times 10^{-14}$& $\Delta a_{\mu}$ & $1.97 \times 10^{-9}$ \\
  \end{tabular}
  \caption{Same as Table~\ref{tab:4} but for scenario {\bf BM II}.\label{tab:8}}
  \end{center}
\end{table*}

For our second representative benchmark scenario, denoted as \textbf{BM II}, we consider a new physics setup characterised by light VLLs that can be copiously produced at hadronic colliders and that predominantly decay into tau leptons. This scenario has been carefully selected to satisfy all constraints imposed during the parameter scan, the corresponding parameters being provided in Table~\ref{tab:8} alongside the values of relevant observables. One of the distinctive collider signatures associated with \textbf{BM II} arises from the pair production and subsequent decay of the $e_7^-$ state. The corresponding signature manifests itself through final states comprising multiple leptons, including at least two tau leptons given that $\mathrm{BR}(e_7\to\tau Z) = 28.71\%$,
\begin{equation}\bsp
  p\,\, p \ \ \rightarrow& \ \  e^-_7\,\,e^+_7 + X \ \ \rightarrow\ \   \tau^-\,Z \, \tau^+\,\,Z + X \ \ \\
  &\quad \ \ \rightarrow \tau^- \ell^+\ell^-\, \tau^- \ell^{\prime+}\ell^{\prime-} + X\,.
\esp\label{BMII_process}\end{equation}
In our analysis, we consider leptonic tau decays together with $Z$-boson decays in electrons and muons. This yields a signal rate of 2.8~fb at NLO+NNLL, as well as a final state exhibiting missing transverse energy. As shown below, the presence of these specific characteristics provides valuable handles on the model. 

\begin{figure*}
  \centering 
  \includegraphics[width=.49\textwidth, trim={1cm 13.95cm 19.5cm 12.2cm},clip]{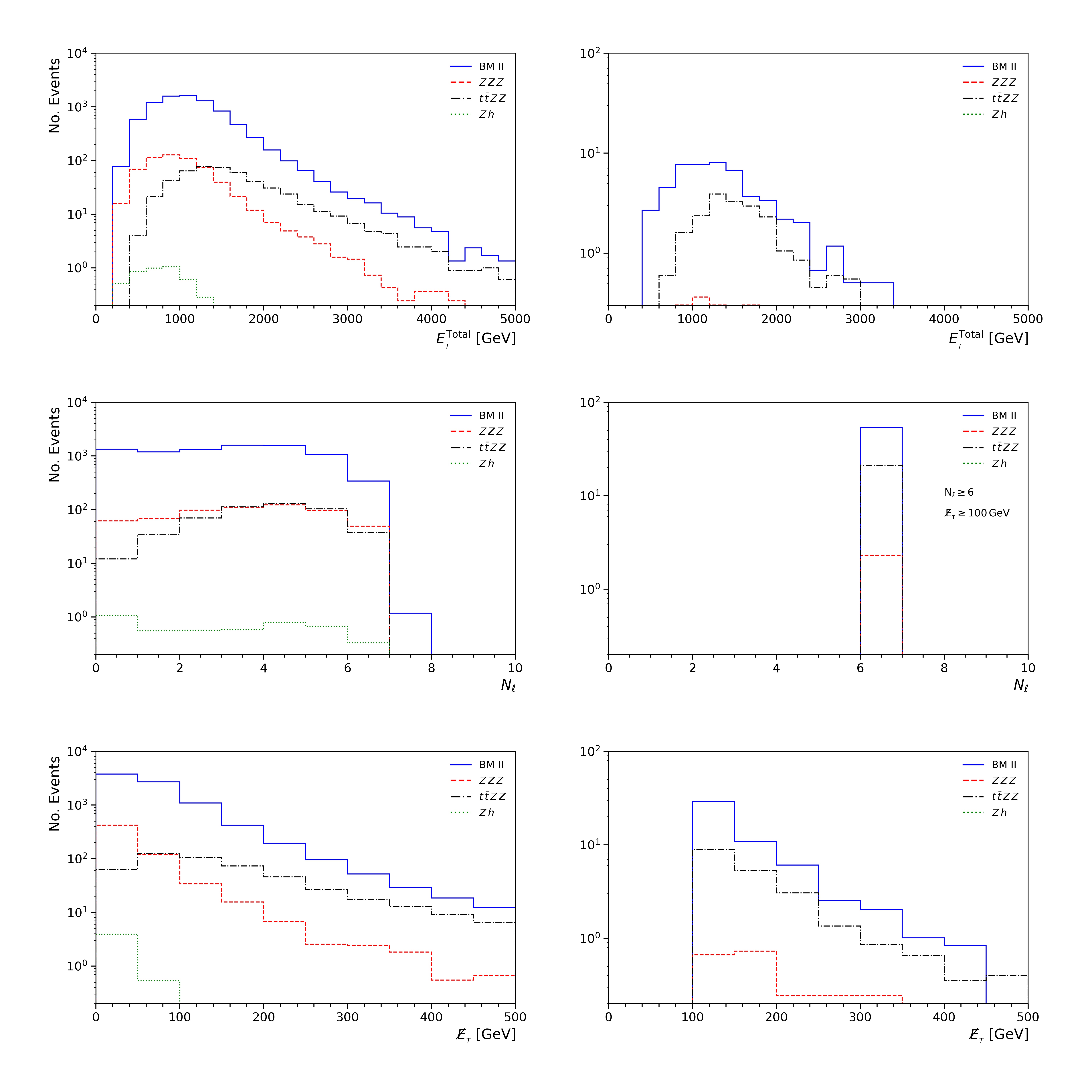}  
  \hfill
  \includegraphics[width=.49\textwidth, trim={1cm 1.75cm 19.5cm 24.4cm},clip]{figures/ppe7e7barBMII}
  \caption{Key observables aimed at distinguishing the \textbf{BM II} signal from the SM background. We display the distribution of the number of reconstructed leptons $N_\ell$ (left), and of the total missing transverse energy $\slashed{E}_T$ (right). Histograms are normalised to a luminosity of $3\,\text{ab}^{-1}$ of $pp$ collisions at $\sqrt{s} = 100\,\text{TeV}$.} \label{fig:11}
\end{figure*}

Backgrounds are similar to those depicted in section~\ref{subsec:BM1}, with cross sections given by $\sigma_{Zh} = 1.52~\text{ab}$, $\sigma_{ZZZ} = 0.2~\text{fb}$ and $\sigma_{t\bar{t}ZZ} = 0.17~\text{fb}$ after including relevant weak boson and tau lepton branching ratios. To extract the signal, events with at least six electrons or muons produced in association with missing transverse energy $\slashed{E}_T > 100$~GeV are selected, with lepton and jet collections defined like in section~\ref{subsec:BM1}. Such choices are illustrated by the signal and background distributions presented in figure~\ref{fig:11}. These basic criteria yield the following number of signal events and background events,
\begin{equation}
  n_s = 53.3
  \qquad\text{and}\qquad
  n_b = 23.8
\end{equation}
with
\begin{equation}
  n_{Zh} \simeq 0,\ \
  n_{ZZZ} = 2.3 \ \ \text{and}\ \
  n_{t\bar{t}ZZ} = 21.5\,,
\end{equation}
and the corresponding sensitivities, considering a 20\% level of systematics on the background, are found to be
\begin{equation}
  s\,=\,7.8\,\sigma\qquad\text{and}\qquad Z_A\,=\,10.3\,\sigma\,.
\end{equation}
Once again, these results demonstrate the clear potential of our setup to distinguish the signal from the background, exploiting a second collider characteristics of the model.

%%%%%%%%%%%%%%%%%%%%%%%%%%%%%%%%%%%%%%%%%%%%%%%%%%%%%%%%%%%%%%%%%%%%%%%%%%%%%
\section{Summary and conclusions}
\label{sec:conclusion}
%%%%%%%%%%%%%%%%%%%%%%%%%%%%%%%%%%%%%%%%%%%%%%%%%%%%%%%%%%%%%%%%%%%%%%%%%%%%%%
In the absence of direct signals of new physics at colliders, indirect constraints play a crucial role in constructing relevant and motivated BSM theories. This often involves reconciling discrepancies between SM predictions and experimental data, such as those observed in the anomalous magnetic moments of the muon and electron that can be further linked to neutrino physics. Our study exploits precisely this. We examined a simplified extension of the SM to which we added three generations of vector-like leptons, each associated with a specific SM family, and hence determined complementary information compared to previous analyses of VLL effects on the leptonic anomalous magnetic moments. In particular, our work includes  the latest experimental constraints on both $\Delta a_\mu$ and $\Delta a_e$, and it also supplements the analysis of magnetic moments and neutrino masses with an extensive collider investigation of the allowed (favorable) parameter regions.

After exhaustively listing possible representations for VLLs and their interaction Lagrangian, we identified parameter space regions compatible with neutrino data and consistent with measurements of the anomalous magnetic moment of the electron and the muon. We determined that this requires, minimally,  the inclusion of two weak VLL singlets (one with the quantum numbers of a right-handed electron and the other with the quantum numbers of a right-handed neutrino), along with a weak VLL doublet with hypercharge $-1/2$. The compatible parameter space regions imply distinct collider signatures, prominently featuring six hard leptons accompanied by either significant total transverse energy or missing transverse energy. We performed a detailed phenomenological analysis to demonstrate the sensitivity of future proton-proton colliders operating at a centre-of-mass energy of 100~TeV to these signatures, focusing on two representative benchmark scenarios with VLL masses in the range of $100-400$~GeV. Our findings underscore that if new physics involves VLLs with properties explaining neutrino data and reconciling discrepancies in anomalous magnetic moments of the light-generation leptons, it will yield signals that will be unmistakable at future proton-proton colliders currently under discussion within the high-energy physics community.

%%%%%%%%%%%%%%%%%%%%%%%%%%%%%%%%%%%%%%%%%%%%%%%%%%%%%%
\begin{acknowledgements}
We thank Kivanc Cingiloglu for discussions in the earlier stages of this project, as well as Mark Goodsell, Olivier Mattelaer and Stephen Mrenna for their respective help with the packages {\tt SARAH}, {\tt MadGraph5\_aMC@NLO} and {\tt Pythia}. Our parameter space scan has been conducted on the B\'{e}luga cluster of the Digital Research Alliance of Canada. The work of MF has been partly supported by NSERC through grant number SAP105354, and that of BF by the grant ANR-21-CE31-0013 (project DMwithLLPatLHC) from the French \emph{Agence Nationale de la Recherche}. For the purpose of open access, a CC-BY public copyright license has been applied by the authors to the present document, and will be applied to all subsequent versions up to the one accepted for a publication arising from this submission.
\end{acknowledgements}

%%%%%%%%%%%%%%%%%%%%%%%%%%%%%%%%%%%%%%%%%%%
\bibliographystyle{JHEP}
\bibliography{ref}
\end{document}